\documentclass[12pt]{article}
\usepackage[left=2cm,top=2cm,right=2cm,bottom=2cm,nohead,nofoot]{geometry}
\usepackage[pdftex]{graphicx}
\usepackage{setspace}
\usepackage[square]{natbib}
\usepackage{amssymb}
\usepackage{verbatim}
\usepackage[usenames,dvipsnames]{color}

\newcommand{\imloc}{}
\newcommand{\baseloc}{../../}

\begin{document}

\title{Valued Ties Tell Fewer Lies: Why Not To Dichotomize Network Edges With Thresholds\footnote{Previous versions carried the title ``The Thresholding Problem: Uncertainties Due To Dichotomization of Valued Ties''.}}
\author{Andrew C. Thomas\thanks{Visiting Assistant Professor, Department of Statistics, Carnegie Mellon University. Corresponding author: email \href{mailto:act@acthomas.ca}{act@acthomas.ca}. This work was supported by grant PO1-AG031093 from the NIA through the Christakis lab at Harvard Medical School and DARPA grant 21845-1-1130102 through the CMU Statistics Department. Thanks to attendees at the SAMSI Complex Networks Workshop, the CMU CASOS Network Science Group and the RAND Statistics Group for comments on earlier editions of this work.} \and Joseph K. Blitzstein\thanks{Assistant Professor, Department of Statistics, Harvard University.}}
\date{\today}

\maketitle

\begin{abstract}
In order to conduct analyses of networked systems where connections between individuals take on a range of values -- counts, continuous strengths or ordinal rankings -- a common technique is to dichotomize the data according to their positions with respect to a threshold value. However, there are two issues to consider: how the results of the analysis depend on the choice of threshold, and what role the presence of noise has on a system with respect to a fixed threshold value. We show that while there are principled criteria of keeping information from the valued graph in the dichotomized version, they produce such a wide range of binary graphs that only a fraction of the relevant information will be kept. Additionally, while dichotomization of predictors in linear models has a known asymptotic efficiency loss, the same process applied to network edges in a time series model will lead to an efficiency loss that grows larger as the network increases in size.
\end{abstract} 


\onehalfspacing

\section{Introduction}

As the majority of publication in relational data and complex networks has derived from the graph-theoretic framework, nearly all of the supporting analytical tools that have been developed are meant to handle binary data input. As a result, there has been a strong tendency towards the transformation of valued data into the binary framework in order to conduct an analysis on the ensemble with particular attention to the individual nodes. This is most often accomplished at the analysis stage through the dichotomization procedure: choose a threshold value, set all ties with equal or higher values to equal one, and all lower to equal zero.\footnote{Dichotomization is also known as compression and slicing \citep{scott2000snah} throughout the literature, and across those disciplines that investigate networks. In previous versions of this work, we referred to the procedure as ``thresholding''; we now use this term to refer only to the censoring of tie values below the threshold, and not the final dichotomization.}

The tendency to dichotomize has been abetted by the simplicity of working with binary outcomes, as well as the visualization methods currently available for graphs. A threshold may also be chosen for the sake of parsimony of analysis, since examining only strong ties simplifies their role in the system under study. Additionally, limiting an analysis to the strongest ties that hold a network together is perceived to be a mechanism for reducing noise that is seen to be caused by a larger number of weaker connections. This is certainly true when using standard algorithms for plotting a network; an excess of ties on the printed page, with respect to the nodes of the system, will obscure other ties that may have more meaning for the dynamics of the system.

If the goal of dichotomization is to learn about the underlying valued system, the outcome may be difficult to quantify. In particular, if the goal is to choose a cut-point that is in some sense ``optimal'', the method for choosing the cut should reflect a minimum loss of information from the valued system. But if the quantity of interest is not meaningful in the valued case -- for example, that all edges are non-zero but most are very small, so that the edge-count diameter of the system is 1 -- then it may prove difficult to choose a condition for optimality.

\begin{figure}
\begin{center}
$\begin{array}{cc}
\includegraphics[width=0.5\linewidth]{\imloc 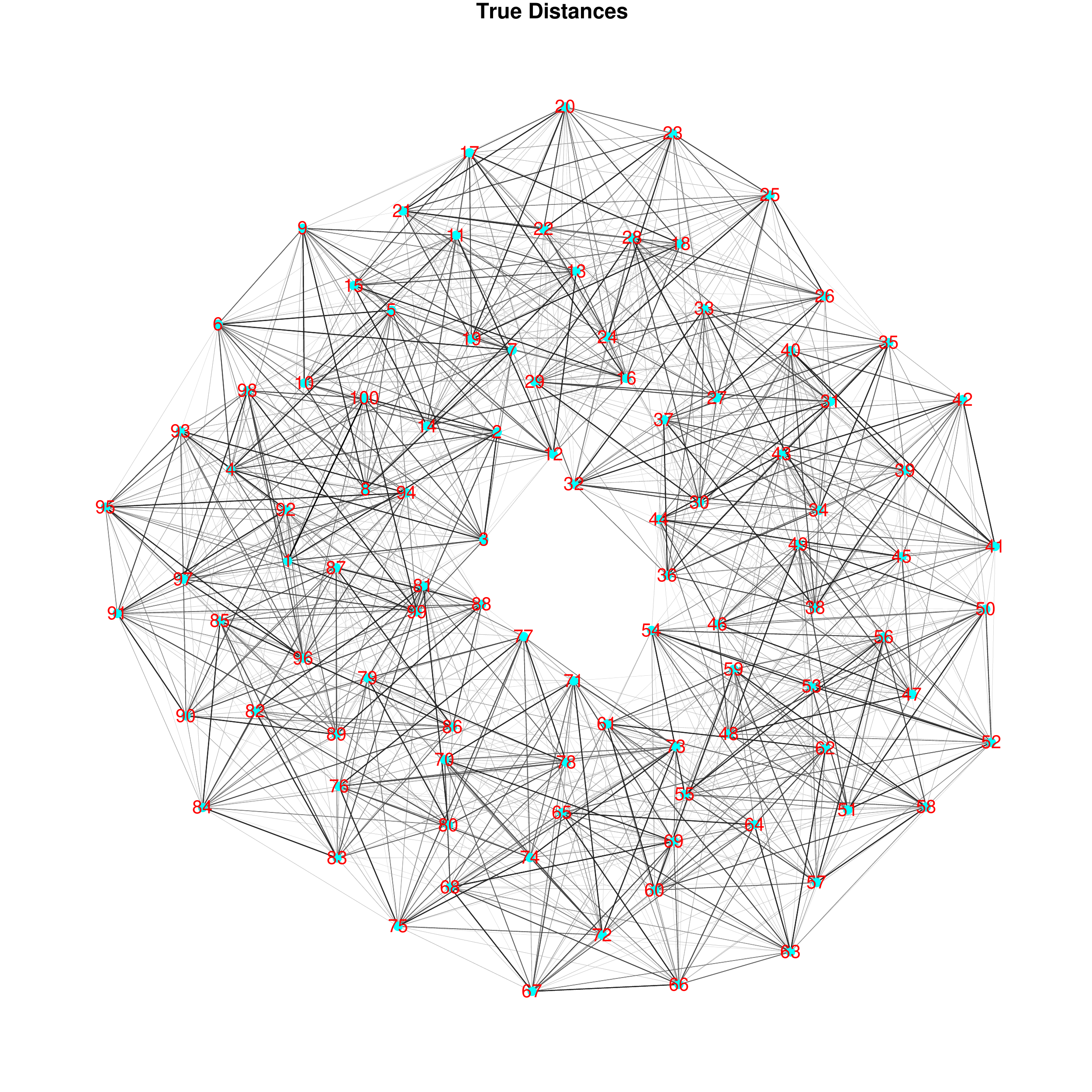} & \includegraphics[width=0.5\linewidth]{\imloc 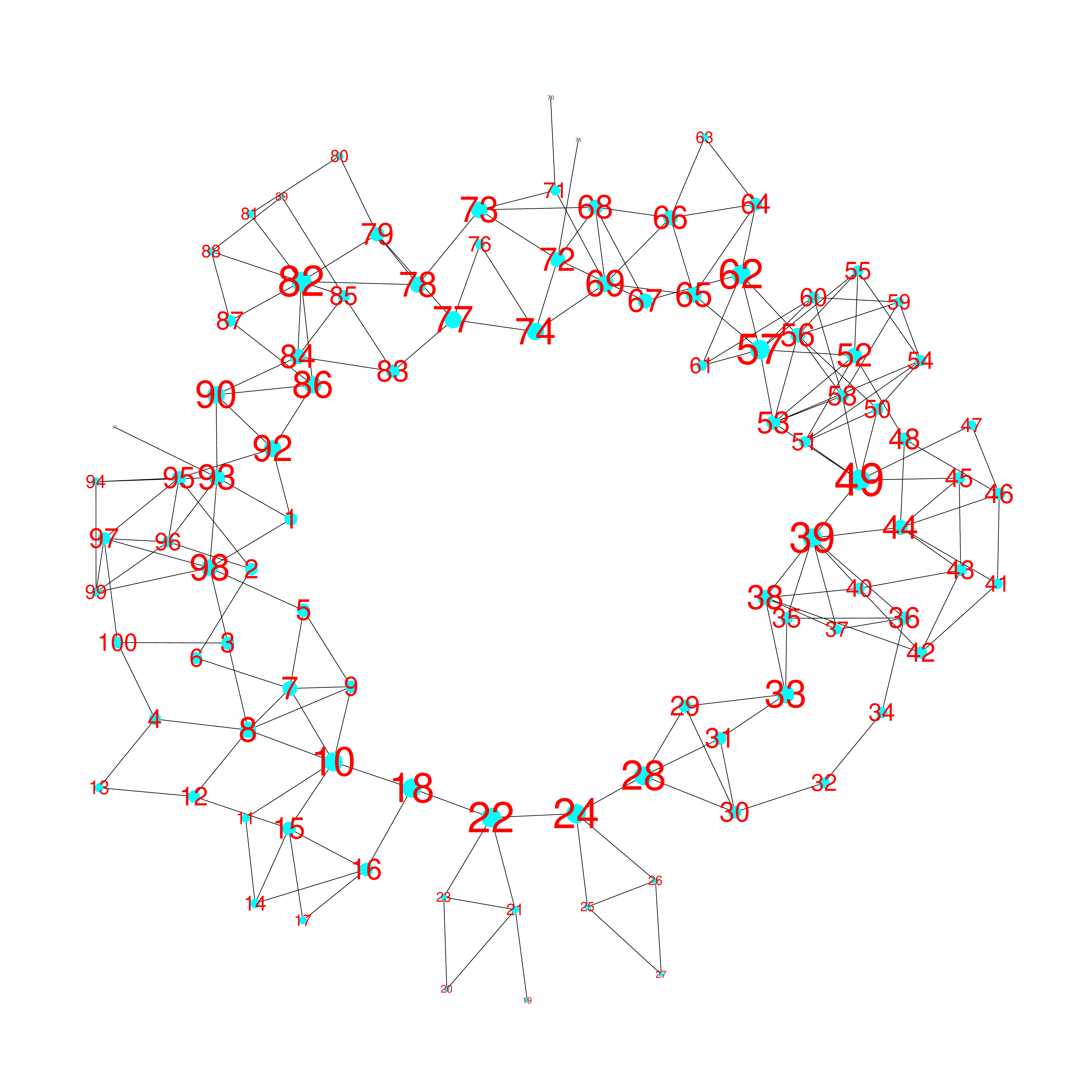} 
\end{array}$
\includegraphics[width=\linewidth]{\imloc 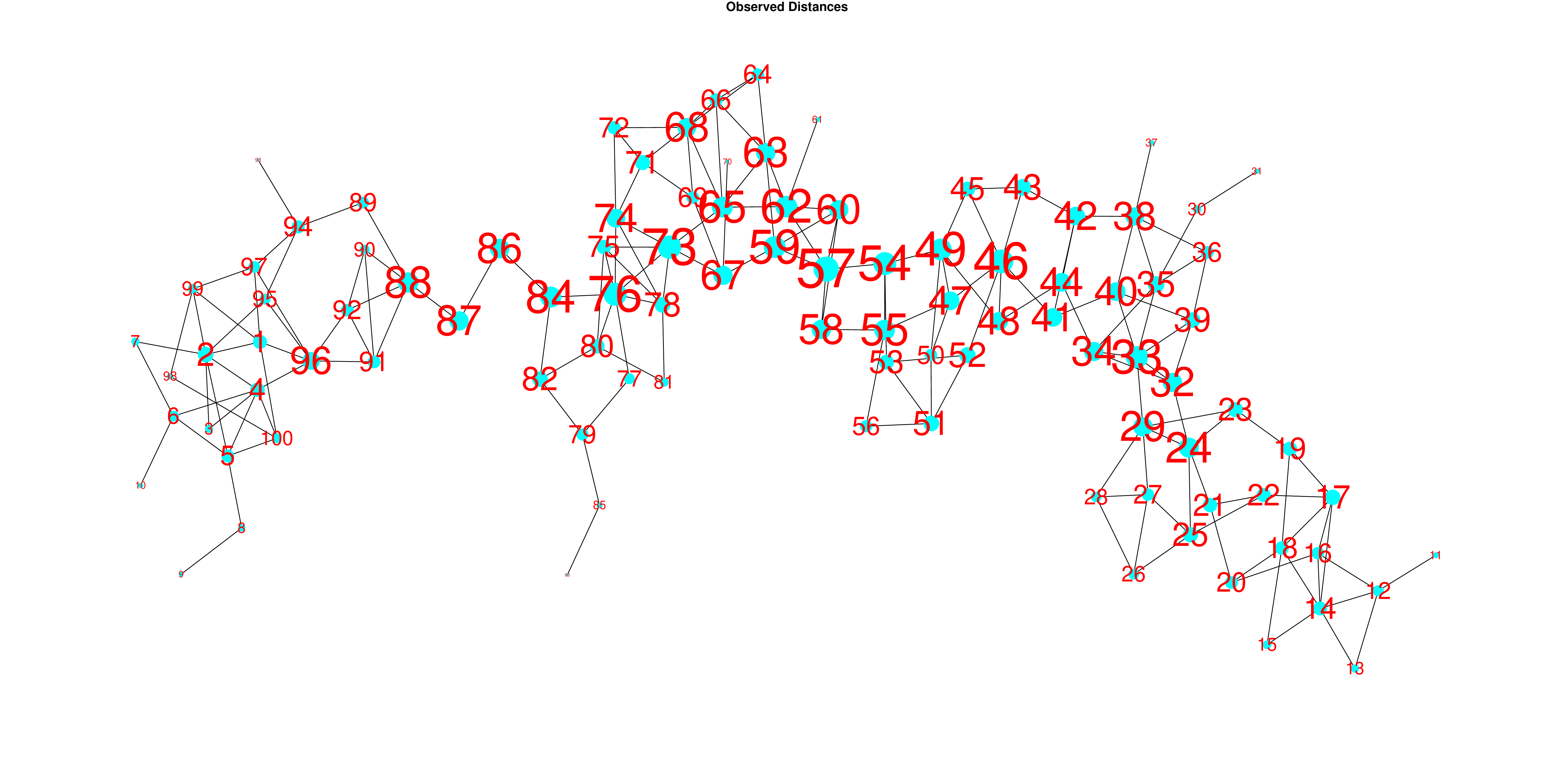}
\end{center}
\caption{\label{worst-case-scenario} A demonstration of when dichotomization can give an extremely misleading picture of an underlying system. Upper left: a 100-node network with ring-type topology and integer-valued ties, produced from a generative model with Poisson-type edge values. Upper right: a dichotomized version of this graph with a cut-point chosen to maintain the giant component. Below: a dichotomized version of a graph from the same generative family with the identical cut-point value. Note that the topological discrepancy, due to the underlying random process governing the strengths of ties, can potentially mislead an investigator on the nature of the connected system. For example, node 18 is located on the ring in the first case, and therefore considerably more central, than in the second case while on the periphery.}
\end{figure}

There are many different classes of input data in the literature that are subjected to dichotomization:

\begin{itemize}

\item \textbf{Correlation or partial correlation} as evidence of network ties (\citet{achard2006rls}, Section 7.3.2 in \citet{kolaczyk2009sand}, \citet{hidalgo2009dnafshp}). Network ties are elicited by measuring the outcomes between two nodes over time or repetitions, and the strength of correlation determines the existence of an underlying tie.

\item \textbf{Count-incidence data} (such as the EIES message data in \citet{freeman1980scsesng}), where a connection value is the number of times two individuals are counted together, be it communication, collaboration or attendance. More classes of count data involving directed transactions are found in the agriculture literature \citep{ortizpelaez2006usn, robinson2007era}. \citet{choudhury2010irsnfic} considers the thresholding problem on modern electronic communication data sets and attempts the same type of procedure we endorse, but with the prediction of future behaviour as the optimality criterion.

A special case of count-incidence is the projection of a binary bipartite network, in which there are two classes of nodes that only have ties across groups, not within. One example of this is the network of memberships of individuals in organizations, which can be projected into an organization-only network with tie strength representing the number of common individuals. In the case of a low-density network, it may suffice to set the threshold at 1; however, this may still result in a significant loss of information.

An example of a count model with noise is Figure \ref{worst-case-scenario}. In this case, the chosen threshold is the minimum value that maintains a giant component, and the resulting topology of the binary network is ring-like. However, for another network generated from the same underlying model, that same choice of threshold creates an line-like binary graph, which has very different topological properties -- among others, it has double the diameter. This is meant to illustrate that if the intent of dichotomization is to reduce noise by simplifying the structure of the system, the choice of threshold may have exactly the opposite effect.

\item \textbf{Categorical/ordinal data} on relationship type (the EIES acquaintance data in \citet{freeman1980scsesng}). In this case, the ordinal data represent the strength of an association between two people as reported in qualitative fashion, such as 

\begin{quote}
\{``never heard of them'', ``acquaintances'', ``casual friends'', ``best friends''\}
\end{quote}

\noindent and subsequent analysis is performed by forming two groups.  

\item \textbf{Rank data} such as in \citep{newcomb1961ap}, in which respondents were asked to identify their order of preference for each other member in the study over a period of several months. This is different from the previous cases since the actual underlying relationships can vary greatly; a person in a tight group of 4 likely has very different feelings for their assignment of ``third-best friend'' than someone in a tight group of 3, whose third choice lies outside their immediate social sphere. 

Thresholds can be taken on an individual's preferences alone, or whether two individuals mutually ranked the other highly. (Note: the former is also another case of the degree-censoring problem, discussed in detail in \citet{thomas2010cocisnpeaa}.)

\end{itemize}

We begin with a discussion of various motivations for dichotomizing a valued data set, and then reviewing work on the consequences of dichotomizing data in the linear modelling literature, and under what circumstances it is maximally efficient. Following this background review, the effect of dichotomization on the geometric summaries of the network is discussed, first through the simulation of various families of valued tie strengths under the GLM framework, then on three real examples from the literature. This is followed by an investigation the choice of threshold on a nodal outcome when the tie is a part of a predictor in a linear model, demonstrate a considerable loss of efficiency as compared to the linear valued case in simulated examples, and show that the coverage probabilities for confidence/probability intervals can be considerably distorted. The chapter concludes with a discussion on the use of dichotomization in general, with recommendations regarding its use in network problems. 

\section{Motivations for Dichotomization\label{thres-motiv}}

There are several reasons why the dichotomization procedure is appealing in an investigation aside from convenience and simplicity. Here are several classes of motivation that are of particular interest.

\begin{itemize}

\item \textbf{Use of Exclusively Binary Methods}. Several classes of models have been designed to incorporate binary information directly, including the exponential random graph model (ERGM or p-star; see \citet{wasserman1996lm} for an introduction), whose inputs are often summary statistics of counts of topological features, and the Watts-Strogatz small-world \citep{watts1998cdsn} and Barabasi-Albert preferential attachment models \citep{barabasi1999esrn} whose generative mechanisms are binary in nature, and also enjoy a large record of verification across many disciplines. No less relevant are our perceptions on the degrees of separation between individuals in a connected system; a friend of a friend is a well-defined notion, whereas a low-strength connection one step away may not be easy to compare to the influence of an individual two very short steps away. 

While it may be possible to refine these mechanisms to incorporate weighted data, these modifications have not yet been verified or published and cannot be relied upon in real-world case studies. As a result, the thresholding mechanism is seen as a reasonable way of extracting information from a valued data set for use in empirically verified methods.

\item \textbf{Ease of Input and Data Collection}. The need to classify continuously-valued quantities into a set of discrete groups is widespread throughout all of science and technology, particularly because of the associated need to make clear decisions based on this information, whether or not there is a distinct change in the behaviour of a system at the threshold level. For example, a person is considered obese if their the body-mass index (BMI) exceeds 30 \footnote{Source: WHO website. \href{http://apps.who.int/bmi/index.jsp?introPage=intro_3.html}{http://apps.who.int/bmi/index.jsp?introPage=intro\_3.html}, accessed July 21, 2009.}, though there is no dramatic difference for two otherwise identical people whose BMIs are 29.5 and 30.5; it serves as a useful benchmark for making medical decisions by preserving a large piece of the information.

In this way, the act of dichotomization is more similar to the rounding of decimal places. However, by rounding too early in the operation, the error introduced will have more opportunities to propagate and magnify through an analysis if not properly tracked.

\item \textbf{Ease of Output in Graphical Representations}. The visual appeal of graphs and networks has contributed to much of the field's attention in the past decade. When plotting a graphical structure, with $n$ nodes and ${n \choose 2}$ undirected edges, it can quickly become difficult to visually discover the most relevant nodes or connections. A clever choice of threshold can illuminate which nodes are most central, which connections the most vital.

\item \textbf{Sparsity of Structure}. In data where there are very few natural zeroes (if any), dichotomization provides a way to select for a small number of connections which are thought to be of the greatest importance to the system, or to nominate a number of ties for more in-depth study. This use, in particular, exemplifies the difference between a ``structural'' zero, in which no tie exists, from a ``signal'' zero, in which the transmission along or activity across a known tie is so small as to render it redundant to the operation of the network.

\item \textbf{Binning To Address Nonlinearity and Reduce Noise}. A quantity that appears to have a linear effect on a short range of scale may behave quite differently over larger scales.\footnote{An appropriate quote: ``Money doesn't always buy happiness.  People with ten million dollars are no happier than people with nine million dollars.'' -Hobart Brown} If there is a nonlinear relationship in the data, binning the data into distinct ordinal categories has many advantages, namely, the reduction of total mean-squared error, and a corresponding increase in power for detecting a true non-zero relationship over an improperly specified linear analysis. By restricting the number of bins to two, the investigator may be imposing a stricter condition on the data than is necessary.

\end{itemize}

Each case has its merits. The first is indisputable, in that binary-valued methods require binary-valued input. When dichotomization is conducted at the analysis stage, the data collection question is moot; however, if done at the design stage, as in a survey or sampling study, the impact is done and the effect must be considered in the analysis. It is difficult to measure the appeal of a graphical display in quantitative terms, though there are visual characteristics that may be apparent in statistical summaries that can be discovered. As for binning, it is one of a large number of data transformation methods that can be performed to address nonlinearity and noise, and a subset of general categorization, which is beyond the scope of this chapter.

\section{Background on Dichotomization Methods\label{thres-back}}

Dichotomization of network ties is often an ad hoc procedure. By experimenting with various cutoff points and examining the properties of the resulting networks, practitioners may choose an ``appropriate'' cut point that purportedly captures the essence of a network phenomenon.

Here are methods that have been traditionally used to dichotomize data without the need for ad hoc standards.

\subsection{Dichotomization of Predictors in Linear Models}

There are many reasons why statistical practitioners might wish to take a quantitative or categorical variable and dichotomize it as an input for a standard linear regression. Principally is the ease of explanation and interpretability of the difference between a ``high'' and a ``low'' group, primarily for the ease of digestion for a lay audience. It is vitally important to choose an ``optimal'' cut-point based on information provided by the predictor alone; to choose a cut-point that depends on the outcome leads to serious issues, the least of which being the invalidity of the p-value for statistical significance \citep{royston2006dcpmrbi} due to an innate multiple comparison between all possible selections.

It has been noted for decades (see \citet{kelley1939sualgfvti} for a historical example) that by choosing a cut-point at the middle, an investigator is selecting points for analysis near the cut-point that are close together in the predictor but are radically separated as a result of dichotomization. It may make more sense to remove the middle points from the analysis entirely; \citet{gelman2009sp} shows that choosing a trichotomization scheme by keeping those points in the upper and lower thirds for a uniformly distributed covariate will maximize the resulting efficiency while still maintaining interpretability.

Splitting the ties into three groups solves no problems for network statistical measurement or for epidemiology. It may, however, prove to be of some benefit in longitudinal model studies where the predictor is the lagged outcome of a neighbouring node multiplied by the tie strength; this is discussed in greater detail in Section \ref{threshold-linear}.

\subsection{Minimal ``Giant Component'' Methods}

The work of \citet{erdos1960erg} established the conditions under which an Erdos-Renyi random graph would contain a ``giant component'', a subset of nodes that are mutually reachable through their connected edges; namely, that the probability of any particular edge existing multiplied by the number of nodes will tend to be greater than 1. The sudden appearance of a giant component with the adjustment of the allowance threshold has been likened to phase transition changes in matter, as well as percolation conditions on somewhat-regular lattices \citep{callaway2000nrafprg}.

The existence of a giant component in a graph has particular implications in epidemiological contexts; if no such component exists, there can be no transmission along the graph. From this idea, the method of choosing a minimum threshold value for which a giant component emerges. If a network is thought to be minimally connected, it would provide a useful upper bound on the effect of information transmission compared to lower thresholds. 

This cut-point is often taken where the network is appearing to grow at its most rapid rate, meaning that the appearance of some nodes and edges over others may appear to be the product of an underlying noisy process rather than the inclusion of links that are specifically responsible for the connectivity of a system. However, it also acts as a point of maximum discrimination between the full and empty states, a natural type of midpoint between extreme values, and therefore deserves some attention.

This method is also popular for graphical purposes as a way of presenting an uncluttered projection of the system in two dimensions (see \citet{hidalgo2009dnafshp, hidalgo2007pscdn} for examples).

\section{Simulation Models} 

With the definition of Generalized Linear Models for various valued network characteristics, there is a basis for considering the effect of dichotomization, both at the selection of various cut points as well as across various instances of random variation.

\begin{figure}
\begin{center}
\includegraphics[height=\linewidth, width=0.32\linewidth]{\imloc 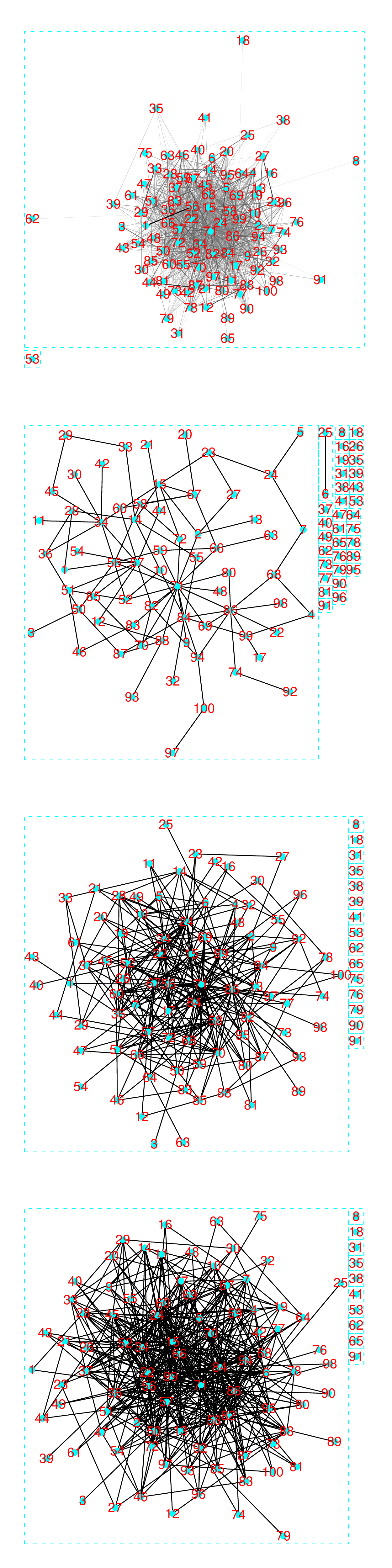}\includegraphics[height=\linewidth, width=0.32\linewidth]{\imloc 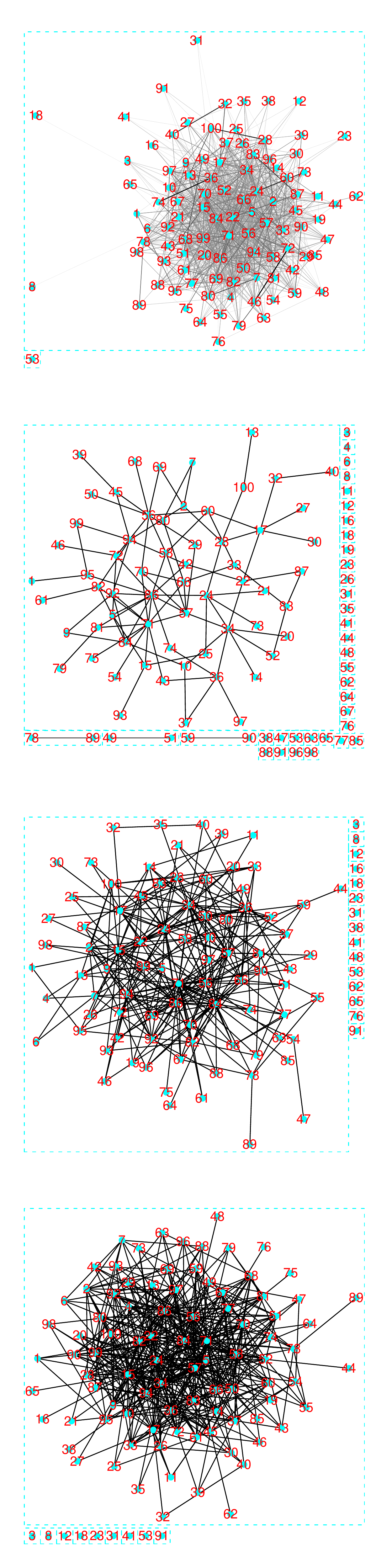}\includegraphics[height=\linewidth, width=0.32\linewidth]{\imloc 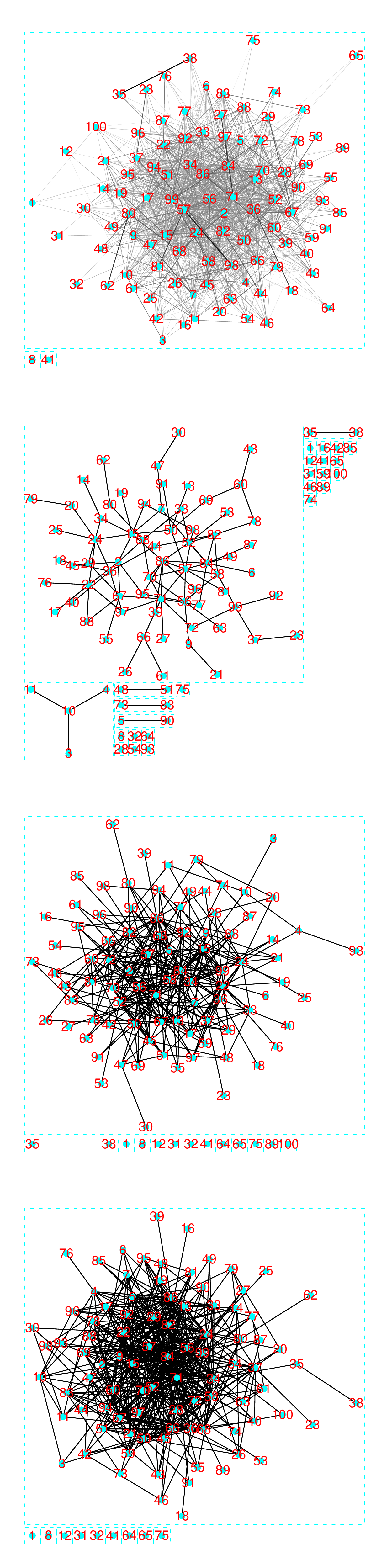} 
\end{center}
\caption{From left to right, three valued networks with the same underlying generative parameters; from top to bottom, the valued graph, plus three dichotomized versions at three different thresholds. Each analysis compares the graphs in each direction, vertical ``within'' analyses for a single valued graph, and averaging over all valued graphs at each threshold value.\label{within-between}}
\end{figure}

The following procedure is used for testing the effect of dichotomization:

\begin{itemize}

\item Select a generative family from the GLM toolkit where edges have nonnegative value:

\[ Y_{ij}|\mu_{ij} \sim \frac{1}{\mu_{ij}}Gamma(\mu^2_{ij}) \]

\noindent and

\[ Y_{ij}|\mu_{ij} \sim Poisson(\mu_{ij}) \]

\noindent are the two generative families used in this experiment. Note that the mean of the Poisson can vary within a single run, leading to the overdispersion that characterizes the heterogeneity present in a Negative Binomial random variable. 

\item Select a series of latent parameters that define $\mu_{ij}$:

\begin{itemize}

\item Sender/receiver effects $\alpha_i \sim N(0,\sigma_{\alpha}^2)$, where a larger $\sigma_{\alpha}$ yields more heterogeneity between nodes.

\item Latent geometric structure: nodes have positions $\vec{d}_i$, and a coefficient of distance vs. connectivity $\gamma$. Nodes can lie equally spaced on a ring of unit radius, or in a single cloud from a bivariate normal distribution. 

\item Latent clusters. Each node is assigned membership in one of three clusters ($a_i=k$), and prefer links either within their cluster or with those nodes in other clusters with propensity $\lambda$. In this simulation, only one of clusters and geometry can be implemented at one time.

\item An assortative mixing factor $\chi$ equal to 0.5, 0 or -0.5 (the disassortative case).

\item The number of nodes in the system.
\end{itemize}

All together, this gives an outcome parameter equal to

\begin{equation} \mu_{ij} = \alpha_i + \alpha_j + \chi \alpha_i \alpha_j - \gamma|\vec{d}_i - \vec{d}_j| + \lambda \mathbb{I}(a_i=a_j); \label{master-sim} \end{equation}

\noindent a list of all options for the above parameters is shown in Table \ref{threshold-sim-types}. 

To keep the parameter values positive, their values are bounded above zero with the transformation function $\mu_{pos} = f(\mu) = exp(\mu-1)\mathbb{I}(\mu<1) + \mu\mathbb{I}(\mu \geq 1)$, rather than setting all negative-parameter draws to zero; in execution, the difference is negligible when threshold values are above 1.

\item Select a ``ladder'' of threshold values, reflecting the changing density and connectivity of the dichotomized systems. These values may be best determined by first considering the average number of edges per node and choosing the threshold that corresponds to that fraction.

\item Given the selected generative model, produce a number of replicates of the valued network (10, for the purposes of this analysis.) For each replicate, create a series of binary networks using the threshold ladder.

\item Given chosen conditions, compare properties of the simulated networks within a single valued instance at all thresholds, taking the average across all instances at each threshold if possible (see Figure \ref{within-between} for an example.)

\end{itemize}

\begin{table}
\begin{center}
\begin{tabular}{c|ccccccc}
Quantity & Values & & & & & & \\
\hline
Nodes & 50 & 100 & 200 & 300 & \textit{400} & \textit{500} & \textit{600} \\
Pop/Greg Signal & 0.1 & 0.5 & 1 & 2.5 & 10 & 100 & \\
Geometry & None & Ring & Cloud & Cluster + & Cluster - & & \\
Geo. Strength & 0.25 & 3 & & & & & \\
Assortative Mixing & 0 & 0.5 & -0.5 & & & & \\
Family & Gamma & Poisson & & & & &
\end{tabular}
\end{center}
\caption{\label{threshold-sim-types} Simulation parameters to investigate the effects of dichotomization in valued networks. For the geometric measures, generated networks have a maximum size of 300; larger networks are implemented in Section \ref{threshold-linear}, the consequences on linear models using dichotomized networks.}
\end{table}

Given the run of these simulations, it remains to be demonstrated how to extract the maximum amount of information from a dichotomized network representation. Estimates obtained through the dichotomized network must have meaning in terms of the individuals within it. In the sections that follow, three levels of effects are examined: static node characteristics, network diameter and dyadic causation. First, the problem of comparing quantities between valued graphs and their dichotomized counterparts in a physically valid fashion is addressed.

\subsection{Valid Estimation Through A Change of Units}

The values and weights in relational data typically have physical meaning. As a result, dichotomization of data is essentially a change of units from an observational measure into a friendship measure, albeit as a lossy many-to-one transformation. Quantities that are calculated under the dichotomized structure will act as valid estimators for the valued quantities if units are accounted for. This is similar to the example set by \citet{gelman2009sp}, in which the efficiency of a predictor is compared between the valued and dichotomized cases. 

While the original measurement has its own scale in terms of a physical quantity (number of communications, minutes in contact, etc.) there is rarely such a definition for the binary tie. In order to distinguish the physical quantity from the model quantity, we define (with an admitted degree of cheek) the unit of a binary social tie, $[binary]$, to be the \textit{Phil}.\footnote{Anthropology defines two blood relations to be ``affine'', suggesting it might be a reasonable unit name as well; however, as  mathematics reserves the word for a class of linear transformation of data, it would simply be too confusing to adopt it in this context.} As an example, consider the electronic messages sent between participants in \citet{freeman1980scsesng} (which is shown in greater detail in Section \ref{example-freeman}), where the unit of interest, $[valued]$, is a message. For a threshold value of 21, all message counts of that value or higher are assigned to be a social tie and set as equal to 1 \textit{Phil}; all below are set to zero. The conversion factor between the two is equal to the difference in means between the high and low groups: 

\[ \frac{[valued]}{[binary]} = \frac{\bar{X}_{high} - \bar{X}_{low}}{1 - 0} = \frac{76.7 - 2.2\,messages}{1\,Phil} = 74.5\,messages/Phil. \]

Since the binary model is constructed to learn something about the valued system that generated it, the estimated quantity can be converted back to the units of the original measurement; even though much information is lost in the transfer, the relative scale between the two remains. In the case of geodesic closeness, the valued and binary equations are each equal to

\[ C_{1/C}(k) = \sum_i \frac{1}{d(k,i)}, \] 

\noindent where $d(k,i)$ is the shortest path from node $k$ to $i$ and have units equal to the original tie strength measure (under the transformation that the path length of a tie is the inverse of its strength) . An estimate of the valued harmonic closeness from the binary is then equal to 

\[ C_{1/C}(k,valued) = C_{1/C}(k,binary)\frac{[valued]}{[binary]} = C_{1/C}(k,binary)*74.5\,messages/Phil. \]

As in the rest of this work, distance is treated as a measure of inverse connectivity. As a result,  calculations of the geodesic shortest-path lengths in the binary case are in units of inverse \textit{Phil}:

\[ d(i,j)^{valued} = d(i,j)^{binary}\frac{[binary]}{[valued]} = d(i,j)^{binary}*1.34 \times 10^{-2} Phil/message. \]

This change of variables will be used when necessary to compute the deviation of the binary-derived estimate from the valued estimate, noting that the choice of threshold, along with the distribution of the valued coefficient, determines the conversion factor.

\subsection{Comparison Measures and Trial Thresholds}

Two options present themselves for choosing an ``optimal'' threshold. One criterion suggests a ``centroid'' threshold value; that is, working only with a set of dichotomized graphs from the same valued graph, the ideal threshold minimizes the sum of rank discrepancies with respect to all others in the set. However, this has an immediate flaw: this measure is sensitive to irrelevant alternatives. For example, consider a series of threshold values that produces a large number of empty graphs. The centroid value will most likely lie in the empty graph set purely due to their multiplicity. Even cleverly constructed alternatives, such as those based on the quantiles of tie strengths in the system, may suffer from this problem if not carefully considered.

Therefore, the only comparisons considered are between the valued graph and each dichotomized version respectively, rather than any comparisons among dichotomized versions, and propose that if any threshold must be chosen, it should be that value that minimizes the deviation of the chosen measure from the original valued graph.


\section{Effects on Geometry: Node Characteristics and Network Diameters\label{threshold-geometry-section}}

This section examines the results of 212 simulations sampled from the proposed space of generative parameters\footnote{This is 212 of a possible 1296 combinations. Additionally, smaller network sizes take far less time to run, leaving 52 of these 212 networks with 200 or 300 nodes, given that larger networks take a considerably longer time to analyze for both geodesic and Ohmic properties.} and compare them according to a series of summary statistics. Each simulation consists of 10 replicates from the underlying structure taken across 30 threshold values, where each threshold is chosen to produce a graph of a specific underlying density. The optimal threshold for each statistic, for each family, is that with the lowest total sum across all 10 replicates.

The statistical measures we consider are based on two families of distance measures on graphs:

\begin{itemize}

\item Geodesic measures, which are based on the shortest path distance $d(i,j)$ between two nodes $i$ and $j$ (see \citet{freeman1979csncc} for an excellent overview of these methods). The reciprocal of this is the closeness $1/d(i,j)$ which has the property that two nodes in separate components have zero closeness, rather than infinite distance.

\item Ohmic measures, which are based on the interpretation of social ties as resistors (or, more appropriately, conductors) in an electrical grid, so that the distance $d_{\Omega}(i,j)$ between two nodes $i$ and $j$ is equivalent to the resistance of the circuit formed by connecting nodes $i$ and $j$ (with symbol $1/G_{ij}^{eq}$, so that $G_{ij}^{eq}$ is the social equivalent of electrical conductance). The notion is useful in physical chemistry \citep{klein1993rd, brandes2005cmbcf} but is also finding new uses in complex network analysis due to its connections with random walks and eigenvalue decompositions \citep{newman2005mbcbrw}. \citet{thomas2009ocindac} gives a more thorough analysis of these measures and their comparisons with their geodesic equivalents; what is most relevant is that these are more sensitive to the total length of all paths that connect two points, to which geodesic measures, concerned only with the shortest single path, are largely indifferent.

\end{itemize}

For each network, valued or binary, there is a collection of graph statistics based on geodesic and Ohmic measures that apply to the individuals within. The choice of threshold affects the node statistics both in absolute terms and relative to each other, and the inherent uncertainty in the measurement of tie values suggests that these statistics vary between different iterations at the same threshold level, hence the increased reliability of using a number of replicates.

For this analysis, three measures of network centrality are considered:

\begin{itemize}

\item Harmonic geodesic closeness, $C_{1/C}(i) = \sum_j \left( \frac{1}{d(i,j)} + \frac{1}{d(j,i)} \right)$;

\item Ohmic closeness, $C_{\Omega}(i) = \sum_j G_{ij}^{eq}$;

\item Fixed-power Ohmic betweenness, $C_P(i)  = \sum_a \sum_{b \neq a} \frac{1}{\sqrt{G_{ab}^{eq}}}\sum_{j \neq i} I_{ij}^{ab}$, as described in \citet{thomas2009ocindac}\footnote{In brief: for all pairs of nodes $(a,b)$, a fixed power of 1 Watt is applied across the terminals corresponding to the nodes, which have an Ohmic inverse distance $G_{ab}^{eg}$. The measured current through node $i$, $\sum_{j \neq i} I_{ij}^{ab}$ determines the importance of the node to current flow between $a$ and $b$.} (relative rank only)

\end{itemize}

Considering the absolute measures of node characteristics, it is simply a matter of calculating the statistic for each node, at each threshold, within each replicate, and converting the estimate into the units of the valued graph. The optimal threshold for that measure is chosen to be that which gives the lowest squared deviation of the statistic for that starting graph.

It may also be preferable to consider only the relative importance of nodes, thereby removing the concern of a change in units. As a frequently asked question of networked systems is ``Who is the most important individual?'' by some set of criteria, rank-order statistics are a logical choice to measure the change of importance of individuals between two instances of a graph. Since there is also far more interest in the more important individuals (those with rank $R_i$ closer to 1) than the less important ones (with rank $R_i$ closer to $N$), a rank discrepancy statistic of the form

\[ D_{ab} = \frac{1}{N}\sum_i \frac{(R_{ai}-R_{bi})^2}{\sqrt{R_{ai}R_{bi}}} \]

\noindent is used, where $R_{ai}$ and $R_{bi}$ are the ranks of individual $i$ in instances labelled $a$ and $b$. Ties in rank are randomly assorted so that, among other factors, an empty or complete graph is uninformative as to the supremacy of one node over another.

\begin{figure}
\begin{center}
\includegraphics[width=0.92\linewidth]{\imloc 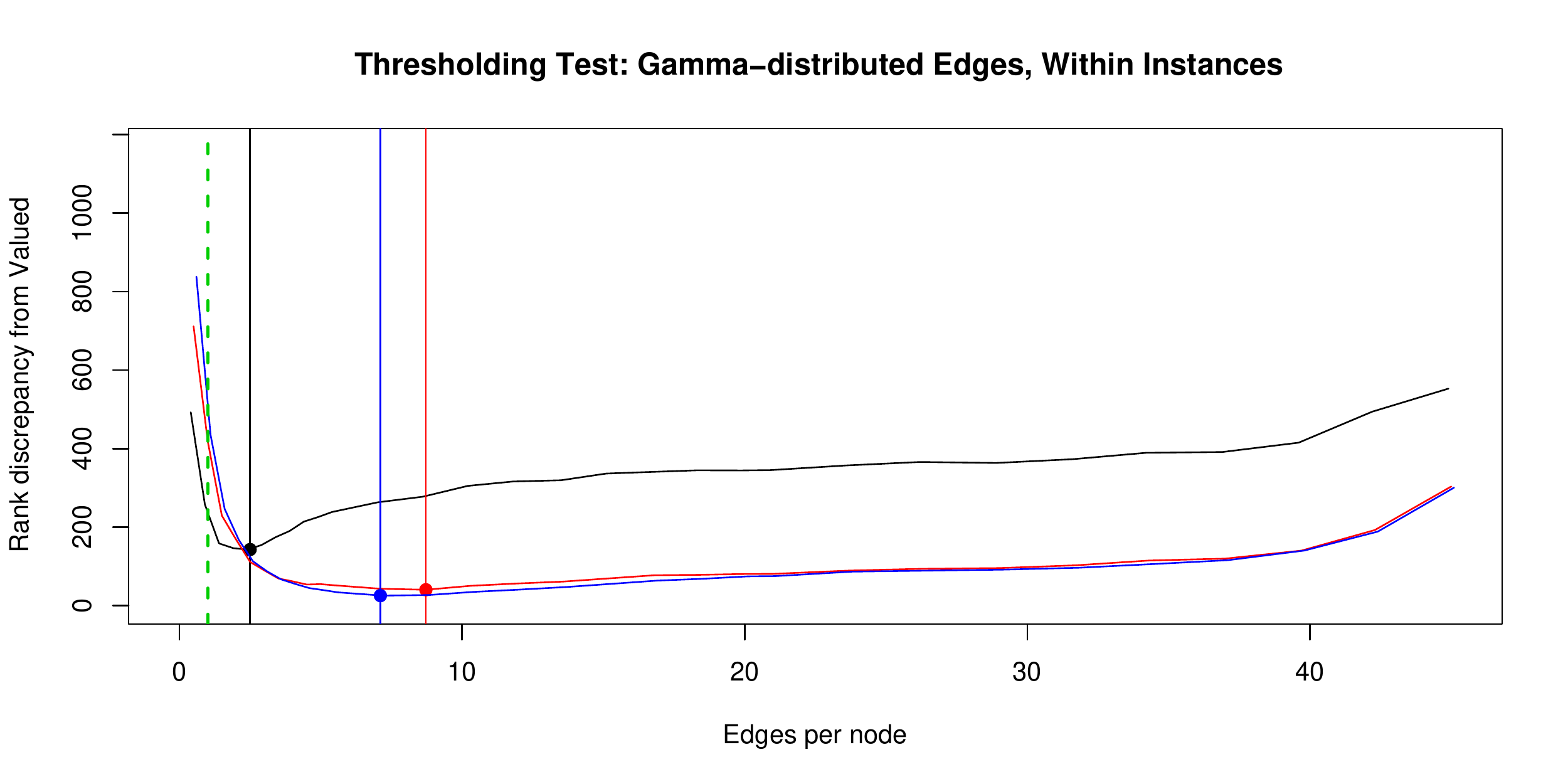}
\end{center}
\caption{\label{threshold-single} The effect of thresholding on importance ranks of individuals for a single generative model class. Each line represents the average rank discrepancy within instances of a valued network from the generative model. Filled circles represent ``ideal'' choices of threshold as compared to the valued case, for harmonic geodesic closeness (black), Ohmic closeness (red) and fixed-power centrality (blue).}
\end{figure}

A single dichotomizing procedure is given in Figure \ref{threshold-single}, for a 50-node network with mild heterogeneity in popularity between individuals and generated by a weak ring structure. The measure of choice is the minimum rank discrepancy between the valued graph and each dichotomized version; these points are highlighted in the figure. All three points are well above one edge per node, the typical point at which a giant component appears (in the case of the theory of \citet{erdos1959rg}).

With the addition of a change in units, a direct value comparison can be made between a dichotomized graph and the original valued model. This brings the values of geodesic and Ohmic closeness into play as fair comparisons. As well, because distances are measured in terms of the inverse unit of friendship, the geodesic and Ohmic diameters for the graph can also be converted from their original values so as to effect a comparison; however, it may prove more sensible to first define inverse geodesic and Ohmic diameter as the minimum non-zero connectivity in a system, so that the units are identical to those for closeness measures (units of \textit{Phil}).

\begin{figure}
\begin{center}
$\begin{array}{cc}
\includegraphics[width=0.45\linewidth]{\imloc 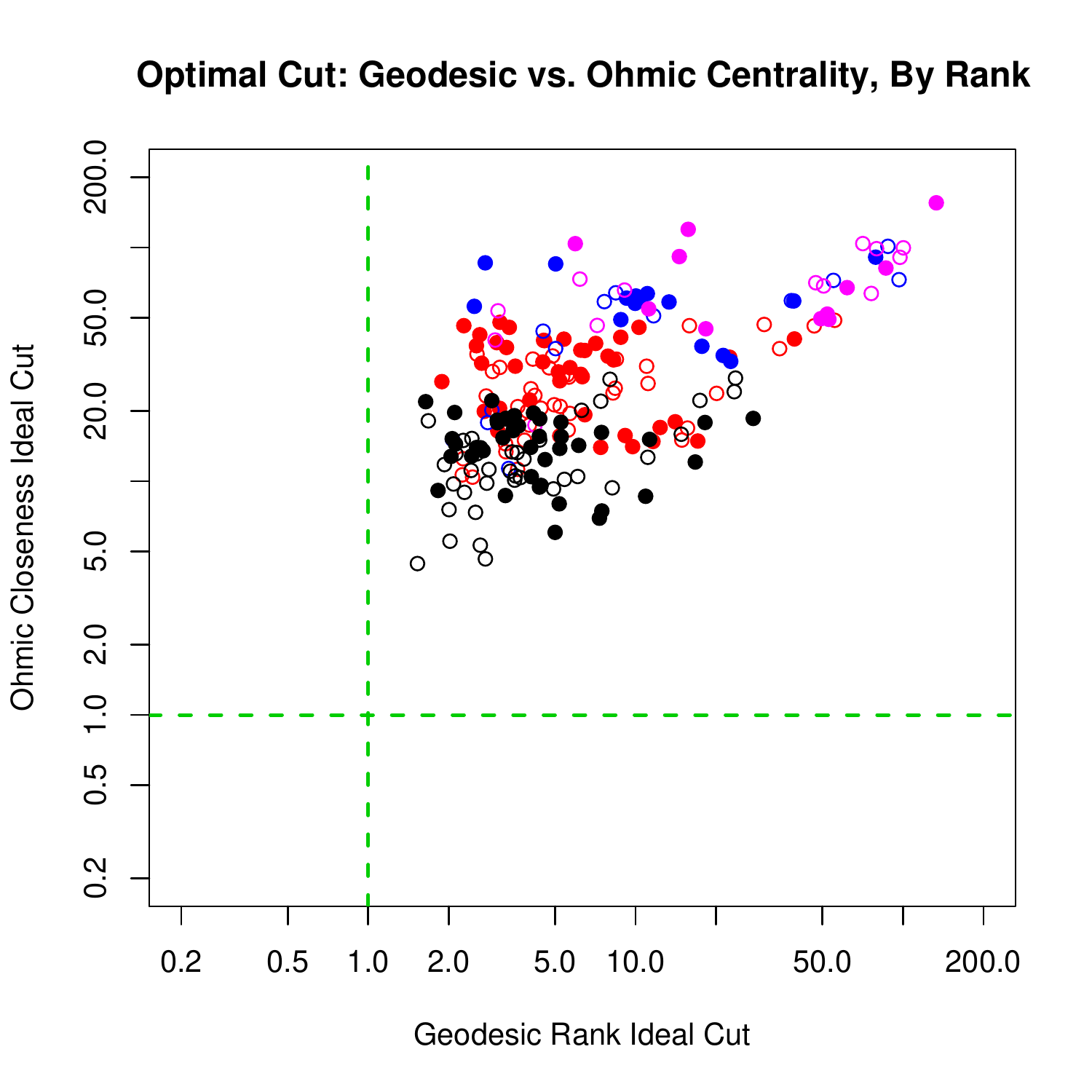} & \includegraphics[width=0.45\linewidth]{\imloc 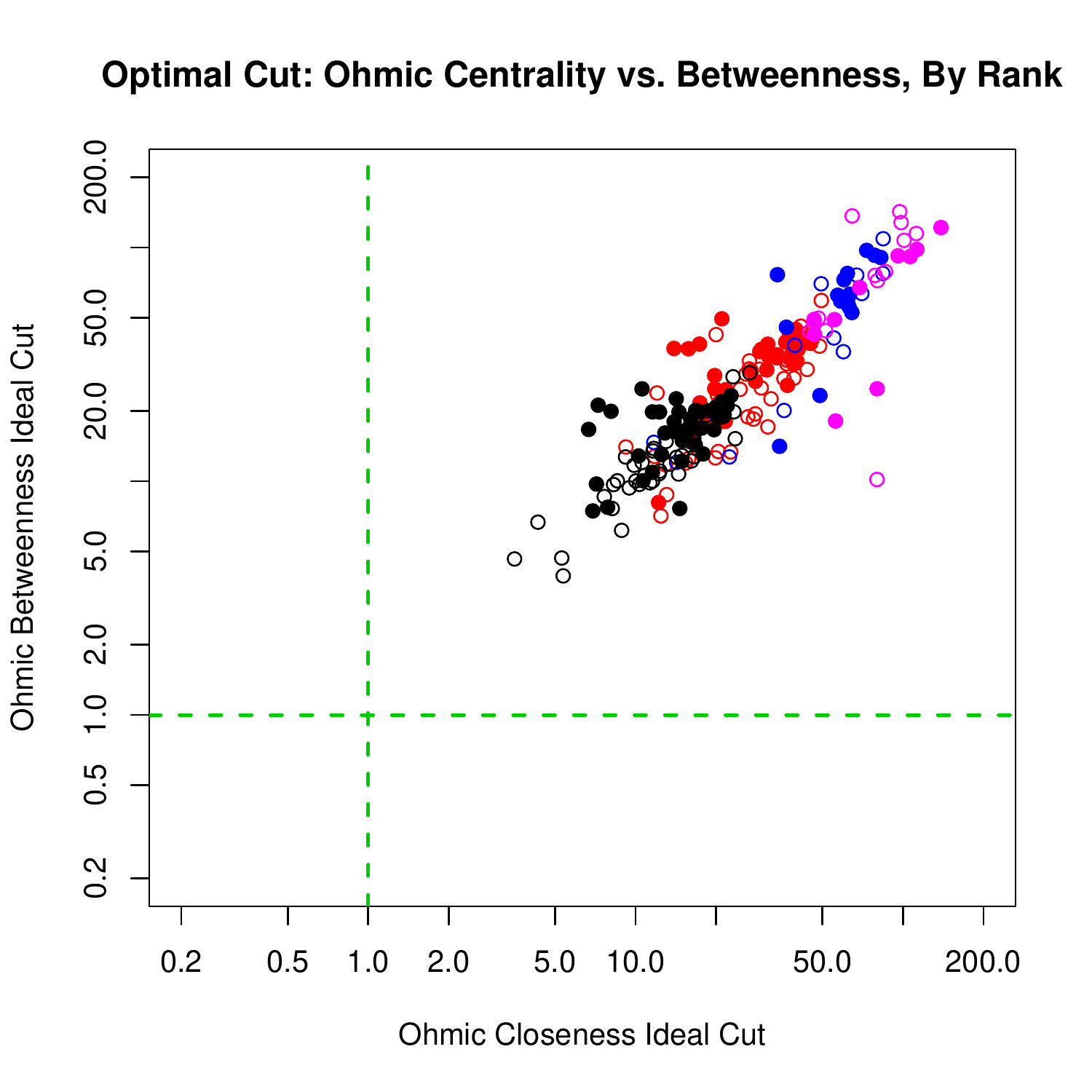} 
\end{array}$
\includegraphics[width=0.55\linewidth]{\imloc 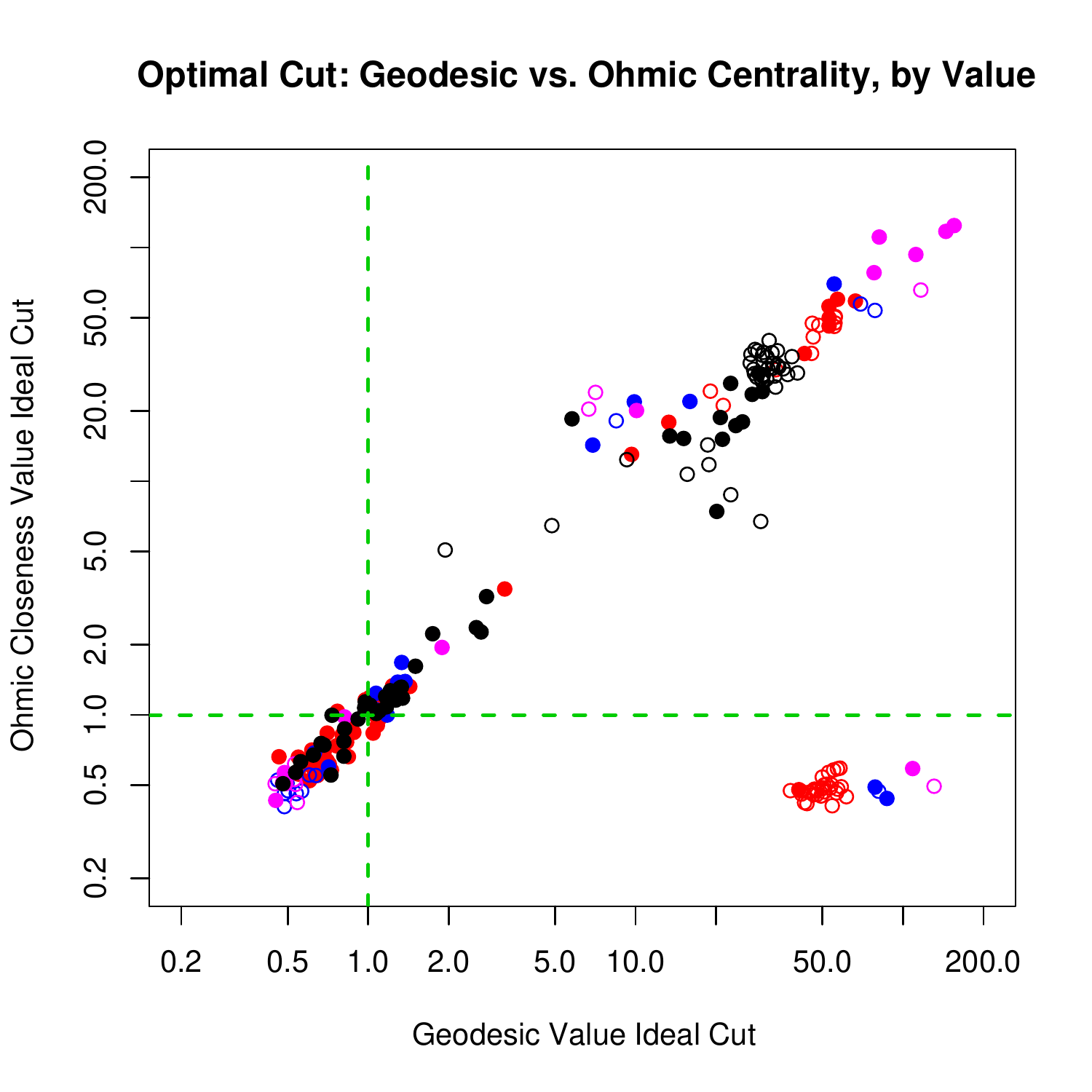}
\end{center}
\caption{\label{thres-scatter} Scatterplot of ideal threshold points based on rank discrepancy statistics for harmonic geodesic centrality, Ohmic closeness and Ohmic betweenness, as well as a comparison of absolute values on closeness statistics. Each point represents a single generative model family; its location is the average edges per node for the optimal threshold across 10 replications. Black, blue, red and pink dots represent simulated networks of size 50, 100, 200 and 300 respectively. For rank-based geodesic or Ohmic statistics, the vast majority of cut points are noticeably above the level of one edge per node, suggesting that the networks produced are far from trivial. This is not necessarily the case of value-based conversions, where the optimal threshold values produce}
\end{figure}

Several scatterplots of results are shown in Figure \ref{thres-scatter}. The first compares geodesic closeness to its Ohmic counterpart, and the differences are apparent. Minimizing the discrepancy for Ohmic closeness requires a higher density binary graph, and hence a lower threshold; this is consistent with the existence of more parallel paths between nodes as an important factor in Ohmic closeness. Additionally, there is a very noticeable effect of network size, such that larger graphs require a higher number of edges per node, but only on Ohmic closeness; if there is an effect for geodesic closeness rank, it is far less pronounced. 

The optimal thresholds by value are a far more unusual story. Many of the ideal points are clustered around 0.5 edges per node, in the region of nearly empty graphs, or roughly one-half the total possible edges per node, in those graphs tending toward full completeness. There are a number that collect at roughly 1 edge per node, the typical minimum for a giant component to appear, but there are very few in the mid-density range of 2 to 5 edges per node, a significantly different result from the rank statistics. Whether this is a consequence of the linearity of the underlying system, or the failure of the unit transformation to properly account for the change in scale, is a subject for later debate; neither are situations that are present in the ranked interpretations, which delinearize the data as a matter of course.

\begin{figure}
\begin{center}
\includegraphics[width=0.55\linewidth]{\imloc 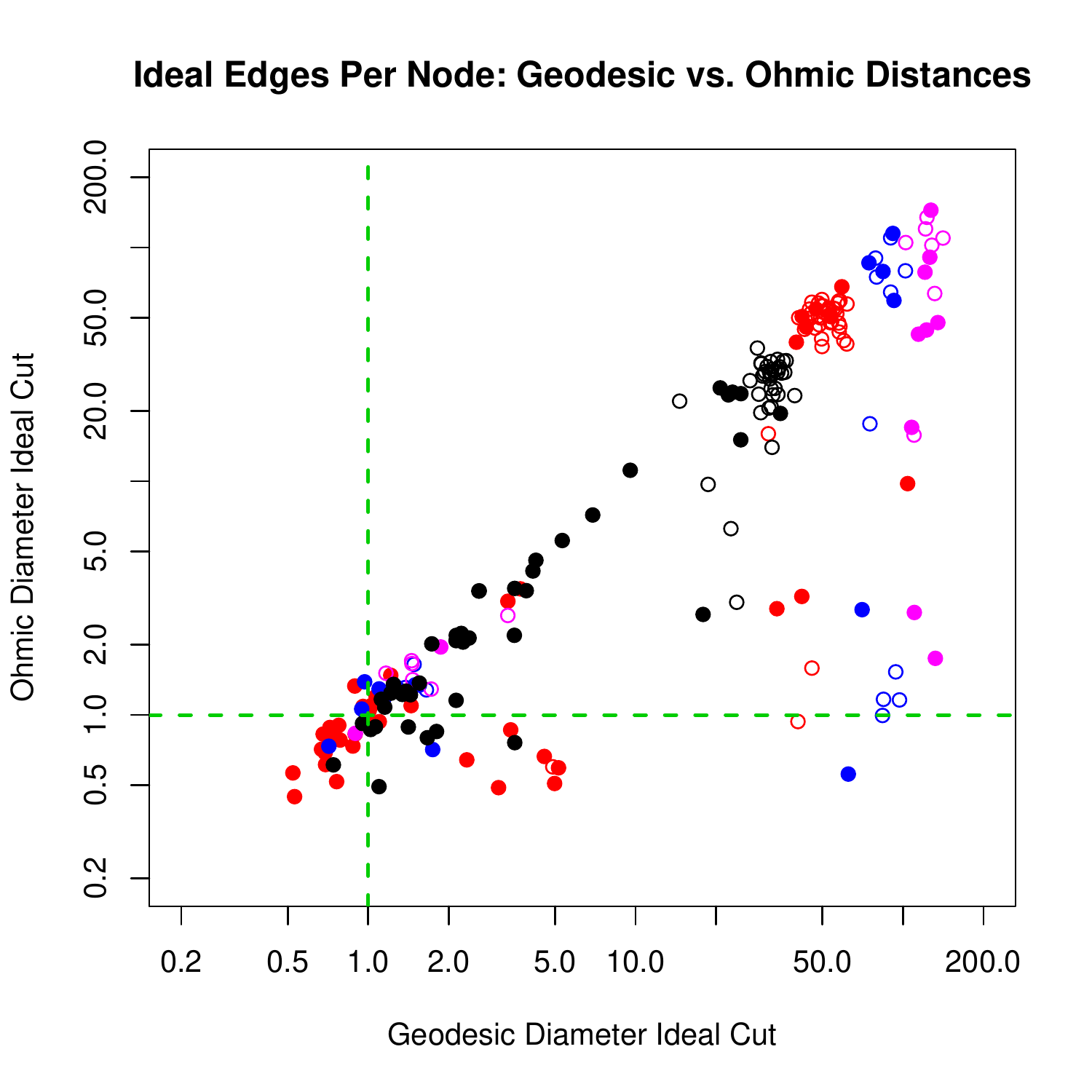}
$\begin{array}{cc}
\includegraphics[width=0.45\linewidth]{\imloc 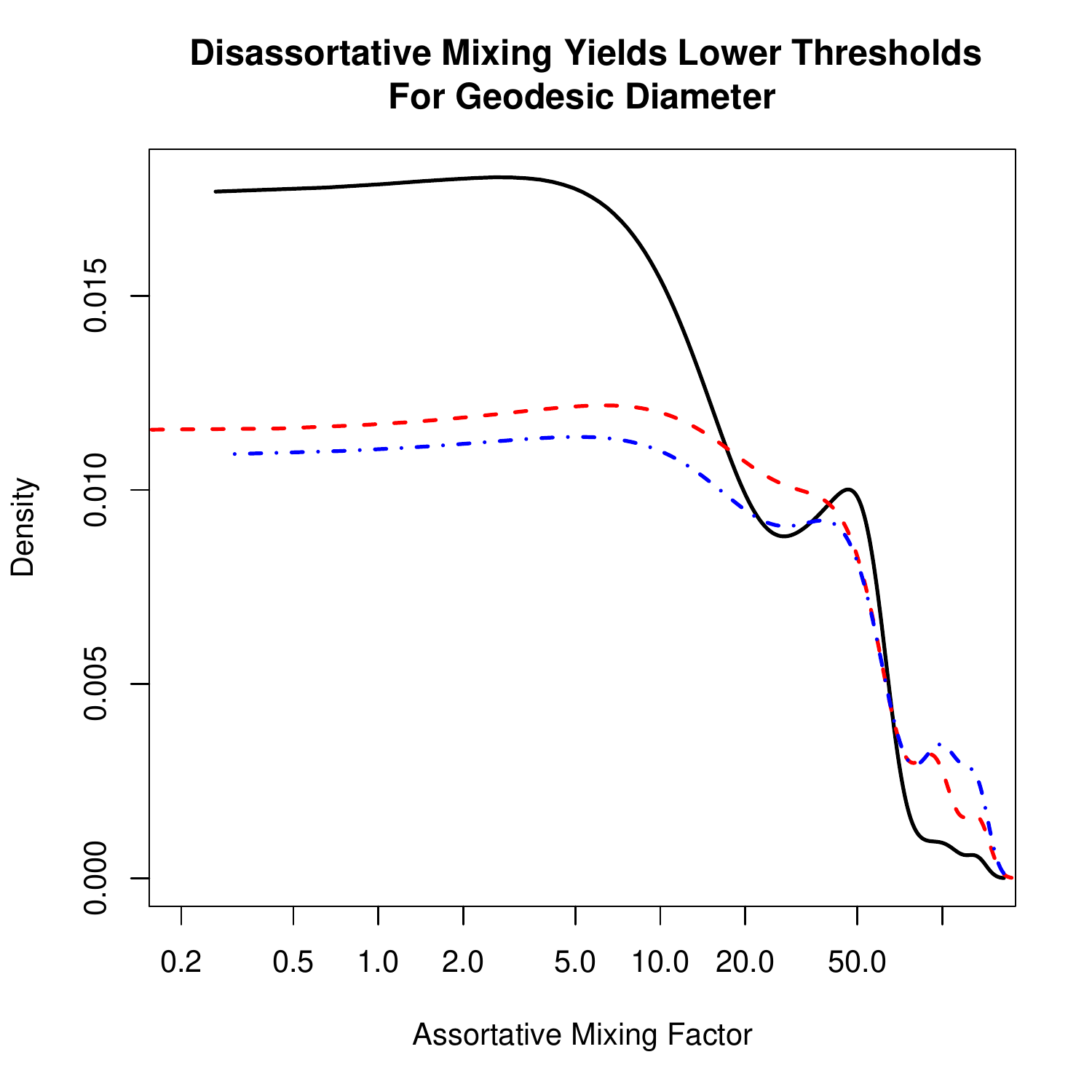} & \includegraphics[width=0.45\linewidth]{\imloc 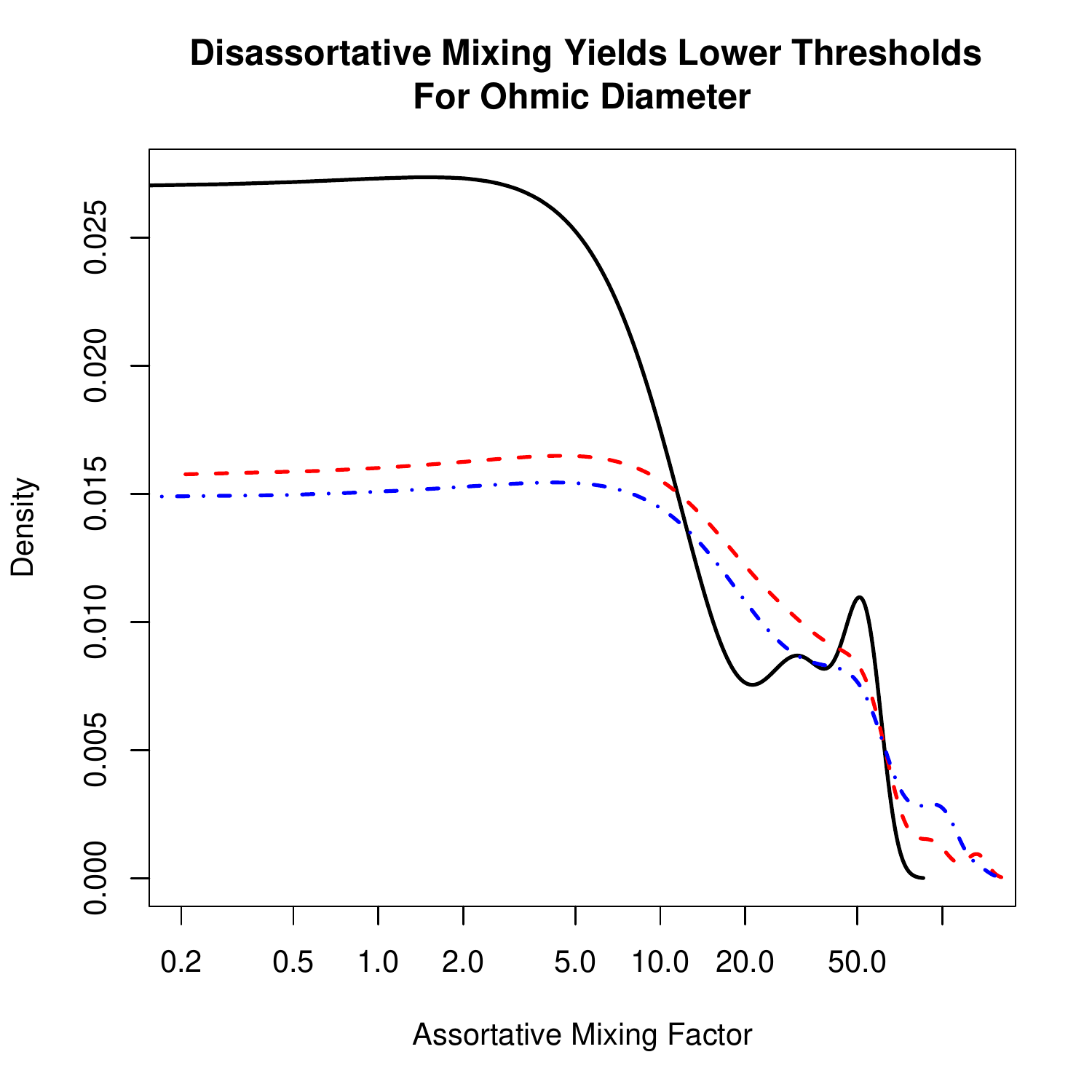} 
\end{array}$
\end{center}
\caption{\label{diameter-disassort} Top: Optimal threshold values for geodesic and Ohmic diameter respectively. Bottom: Kernel density plots for the optimal threshold for geodesic and Ohmic diameter grouped by the assortative mixing constant. The solid black line in each case is for additional disassortative mixing and has higher density for lower threshold values; red and blue represent no adjustment and additional assortative mixing respectively. }
\end{figure}

The same situation is present when examining diameter, though to a lesser extent. Figure \ref{diameter-disassort} summarizes the optimal thresholds for each simulation in terms of geodesic and Ohmic diameter. There is a considerable concentration of points at very sparse and very dense graphs, though there are many more intermediate cutpoints for both geodesic and Ohmic diameter. The cutpoints for Ohmic diameter are often at lower densities, with higher thresholds, than in the geodesic case; this is the opposite of the findings for rank-based statistics for centrality, though the number of points that are in this region is a small fraction of the total. 

\subsection{Results by Generative Parameter}

Each of the generative parameters for the simulations have some impact on the optimal threshold points for one of the statistics of interest. One is the effect of assortative mixing by popularity on diameter, as seen in Figure \ref{diameter-disassort}. In cases where there is significantly strong additional disassortative mixing -- that is, in the case where high-degree nodes are more likely to connect to low-degree nodes -- the required threshold for diameter is considerably higher, so that the number of edges per node is much smaller and a less dense graph is required than in cases with nonnegative assortativity on popularity. This suggests that the base structure for the network is captured by a ``hub and spoke''-type model, so that most nodes are captured with a minimal number of edges, which then form the backbone of the network.

\begin{figure}
\begin{center}
\includegraphics[width=\linewidth]{\imloc 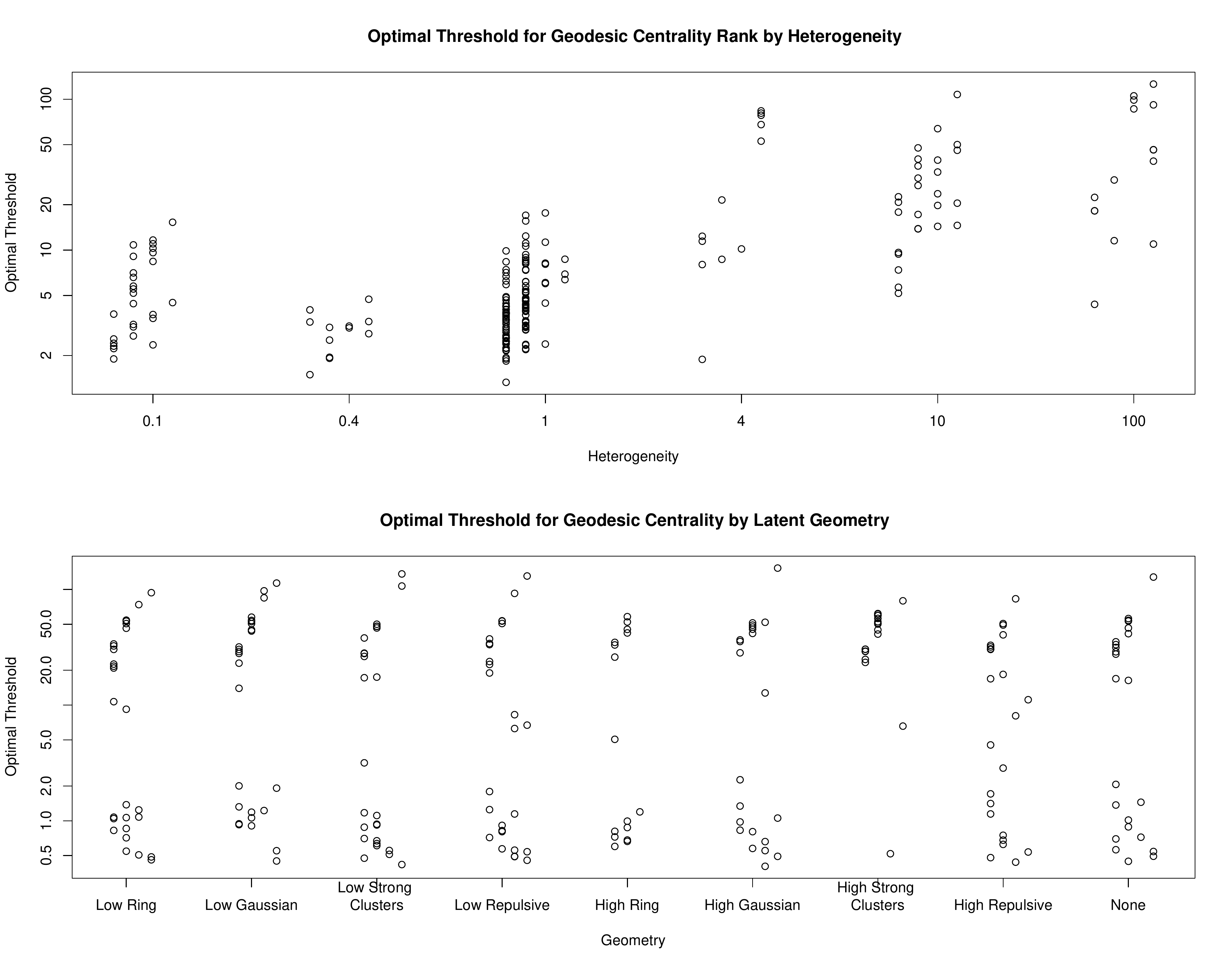}
\end{center}
\caption{\label{threshold-by-coefficient} Two examples of how an input parameter can affect the optimal threshold level; x-coordinates represent various values of the input parameter, while each column within a parameter value represents the network size. Top: Increased heterogeneity in node popularity raises the optimal density for geodesic centrality rank. Bottom: Of the latent geometries proposed, only systems with (relatively) strongly self-connected clusters require high densities to retain geodesic centrality values. }
\end{figure}

Two other effects of generative parameters on the optimal threshold points are demonstrated in Figure \ref{threshold-by-coefficient}. First is the effect of heterogeneity of popularity, or the standard deviation of the underlying parameter $\alpha_i$ in Equation \ref{master-sim}. The optimal threshold rises with the degree of heterogeneity in node popularity, suggesting that in cases of extreme discrimination between node popularity, more ties are needed to accurately represent the valued network in binary terms.

Second is the effect of latent geometry on the optimal threshold. For the nine suggested geometries presented (4 geometries, two levels of effect, plus ``none''), one shows a strong discrepancy from the others: the situation where nodes are located in clusters and show a high preference for ties within their own cluster. In this situation a higher density is necessary, and hence a lower threshold, likely because the ties between clusters tend to be both weaker and essential for the full connectivity of the system (in a situation reminiscent of the ``strength of weak ties'' hypothesis of \citet{granovetter1973swt}).

\section{The Consequences of Dichotomization on the Geometry of Previously Published Examples}

Having simulated and analyzed a wide range of synthetic networks, it is worthwhile to investigate the consequences of dichotomization on three real valued networks.

\subsection{Freeman and Freeman's EIES Communications\label{example-freeman}}

The EIES communications data set \citep{freeman1980scsesng} contains a record of the number of times a group of 32 social network researchers communicated electronically with one another over a period of time, using a precursor of modern email. There is a high degree of heterogeneity in the link values; more than half, 532 of a total of 992, are valued at zero; more than one quarter (258) are valued at more than 10, and 33 are valued at more than 100.

\begin{figure}
\begin{center}
\includegraphics[width=0.95\linewidth]{\imloc 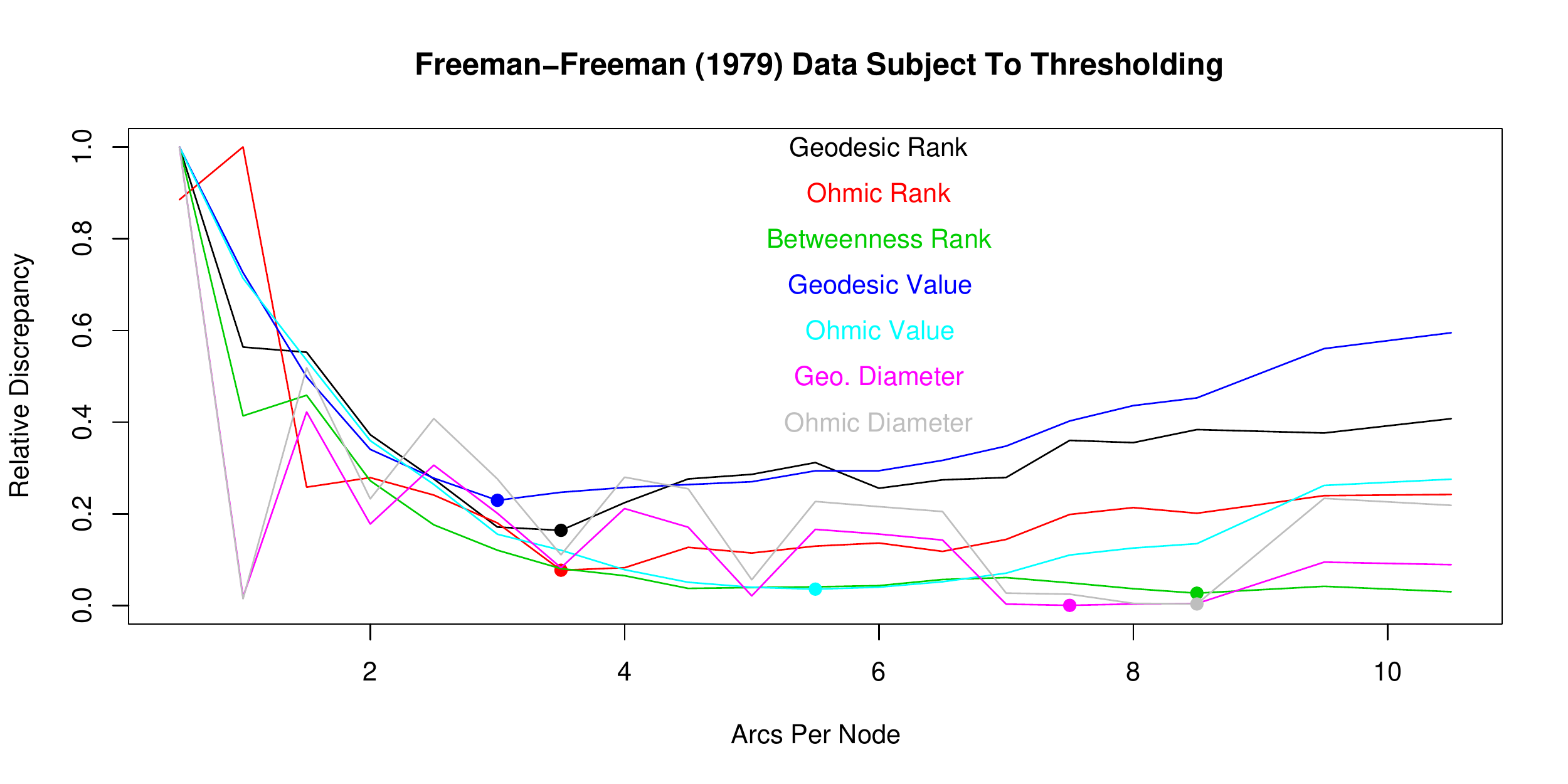}
\end{center}
\caption{\label{freeman-table} For the directed-arc message data set of \citet{freeman1980scsesng}, rank discrepancies and mean-squared errors (relative to the maximum in each case) for the thresholding procedure across seven conditions for 20 threshold values: geodesic, Ohmic and power-betweenness centrality in rank; geodesic and Ohmic centrality in value, and geodesic and Ohmic diameter. There is no clear consensus about which threshold is preferable if any, given the wide range of preferred cutpoints.}
\end{figure}

A full profile of threshold trials can be seen in Figure \ref{freeman-table}. There is no one clearly preferred ratio of arcs to nodes, though a value between 3 and 3.5 is satisfactory to preserve closeness, and a much higher 8.5 arcs per node to preserve betweenness. To minimize the distortion in diameter, the ``best'' estimate is visually bimodal; the global minimum is found at 7.5-8.5 arcs per node, though there is also a significant minimum at 1 arc per node, corresponding to a subgraph of 9 highly prolific people from the total 32.

\subsection{Achard's Brain Wave Correlations\label{example-achard}}

\begin{figure}
\begin{center}
\includegraphics[width=0.95\linewidth]{\imloc 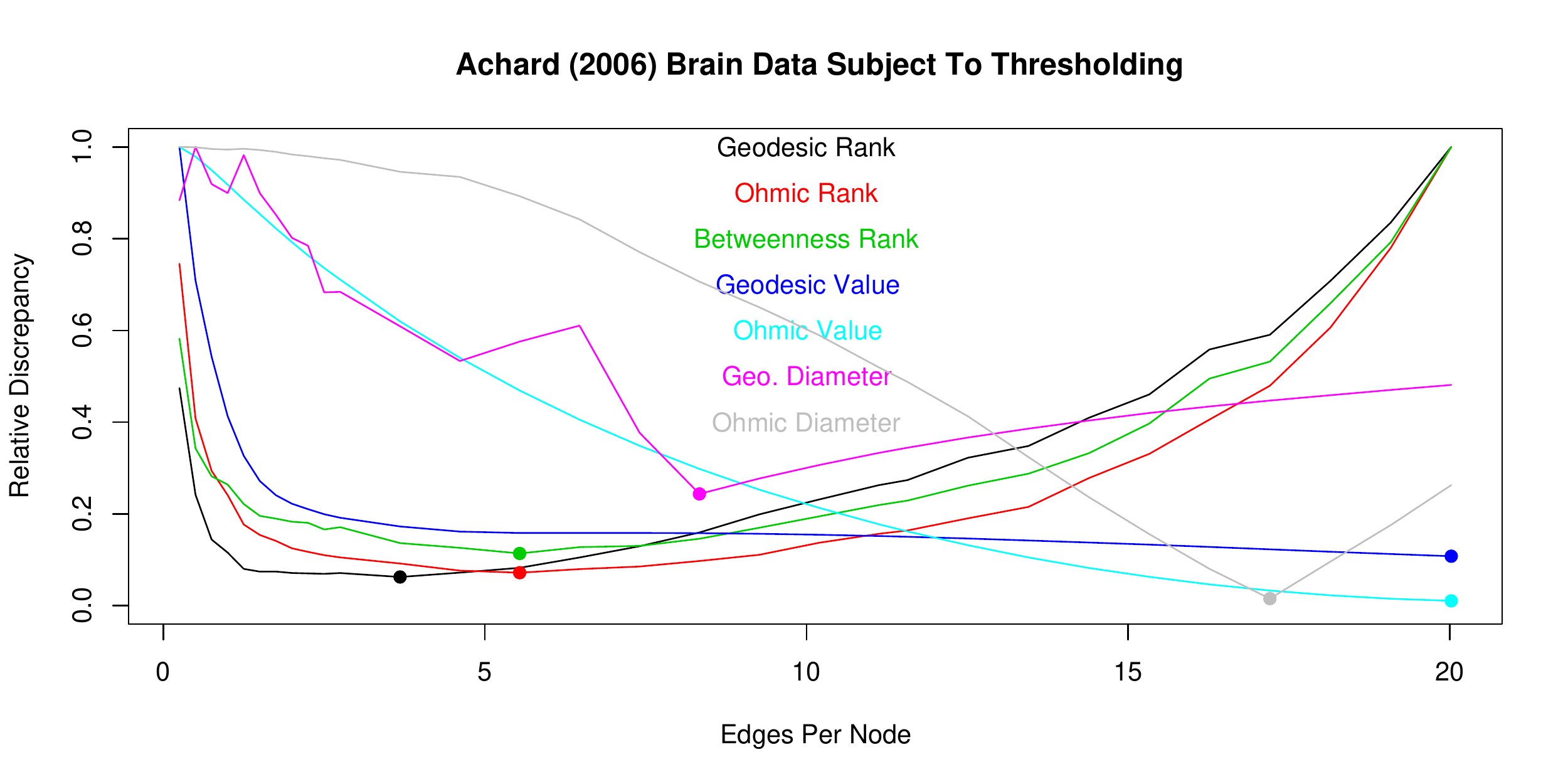}
\end{center}
\caption{\label{achard-table} For a version of the data set of \citet{achard2006rls}, rank discrepancies and mean-squared errors (relative to the maximum in each case) for the thresholding procedure, averaged over 10 generated replicates. Threshold values between 0.22 and 0.26, corresponding to 7 to 11 mean ties per node, produce dichotomized versions that best preserve the relative ranks of the nodes. Accounting for the unit transformation of the ties, a lower threshold appears to preserve distances to a greater degree.}
\end{figure}

\citet{achard2006rls} measure the correlation structure of fMRI data in the brains of various subjects. To produce a network structure, a partial correlation method is used. Several simulated replicates of this set are subjected to dichotomization; Table \ref{achard-table} lists the optimal threshold points for each of the statistical measures considered. It is interesting to note that the cutpoints that preserve relative rank (between 0.22 and 0.26) are far higher than those that preserve distance (one at 0.17, three at 0.056 or fewer). 

\subsection{Newcomb's Fraternity}

The observations of \citet{newcomb1961ap} form a time series at weekly intervals of the mutually ranked preferences for members of a fraternity, previously unacquainted, over 15 weeks. The data are transformed from their original preferences (1 through 16) to fractions ($\frac{16}{16}$, $\frac{15}{16}$, ..., $\frac{1}{16}$) in order to construct a valued scale; no ties were permitted in the original surveys. It is interesting to note that because each row of the sociomatrix contains the same elements and evenly spaced, the conversion factor in each non-trivial case is equal to $1/2$, meaning that there is no length distortion between choices of threshold value.

\begin{figure}
\begin{center}
\includegraphics[width=\linewidth]{\imloc 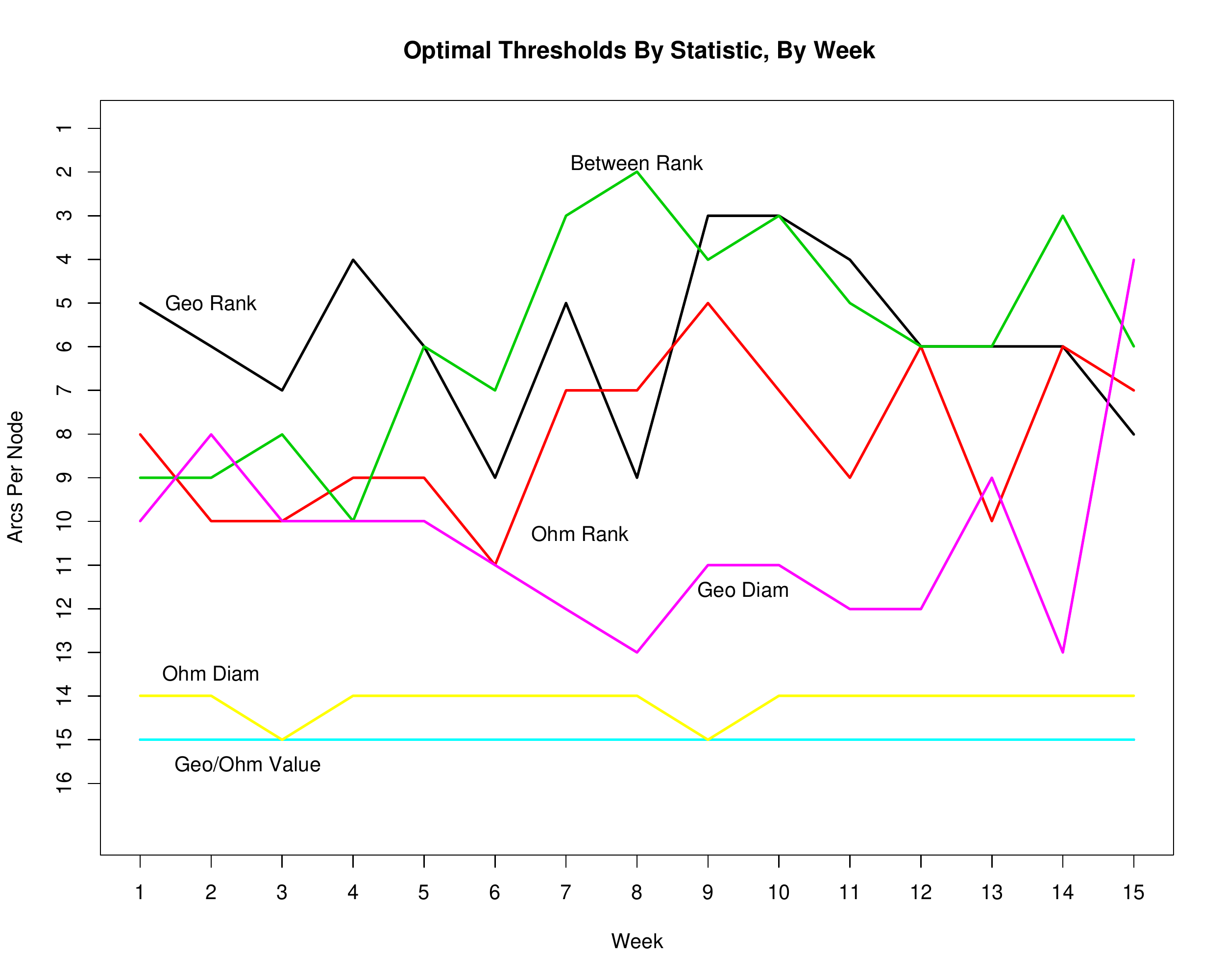}
\end{center}
\caption{\label{newcomb-ideals} The optimal threshold for seven statistics, week by week, in Newcomb's fraternity rank preference data. The values vary across statistics of choice; in particular, optimal thresholds for Ohmic diameter and value-based geodesic and Ohmic centrality are all at levels near to the point where the complete graph appears.}
\end{figure}

The optimal thresholds for the statistics of interest are shown in Figure \ref{newcomb-ideals}. There is considerable variability across the weeks within the statistics, suggesting that a single choice of threshold across all weeks would be suboptimal if the underlying linearity is in fact valid.


\section{\label{threshold-linear} Effects on Tie Coefficients in Linear Models}

One of the prime reasons for modelling systems as networks is the convenience of an explanation for the evolution of a system: phenomena travel between individuals along network ties. This is the essence of the study of social influence on a network. A linear model construction of this influence suggests that a quality possessed by one individual at a particular time will affect the level of the same quality in a neighbour at some later time. 

While there is no need to threshold in the immediate context of a neighbour, there is a particular interest in creating a system where removed degree is a measure of interest. Among other studies, \citet{christakis2007solsno3y} investigate the epidemic properties of obesity in a population by separating the influence of an individual's friends by radius -- that is, radii 1 through 3 represent an immediate friend, a friend-of-a-friend, and a friend-of-a-friend-of-a-friend respectively. By choosing a cut-point for where friendship begins, rather than to construct a distance metric for all individuals, this decomposition can distinguish between various social relationships and establish the effective degree of influence for any one individual.\footnote{Investigations based on the Framingham Heart Study, such as \citet{christakis2007solsno3y}, do not actually use thresholded valued data in their assessments, but instead have censored out-degree for friendship counts due to the construction of the study; see \citet{thomas2010cocisnpeaa} for more information.}

It may, however, prove to be foolish to begin investigating a system with network effects beyond direct connection without examining the simplest cases. Therefore, the remainder of this section is dedicated to one such simple case: the one-step time evolution of a networked system where node behaviour is determined by an autocorrelation term as well as a ``tie'' effect. Following the definition of the model, the full simulation procedure for each model considered is given. Three outcomes are examined: the optimal cut point produced by the model, the relative mean squared error of the thresholded system with respect to the original, and the coverage probabilities of the estimators for the tie coefficient.

\subsection{Setup}

Consider the property of a node at two time points, time 0 ($Y_{i0}$) and time 1 ($Y_{i1}$), or the \textbf{past/present node properties} respectively. A typical linear model setup for measuring the effects of ties on the evolution of node properties takes the form

\begin{equation}
Y_{i1} = \mu + \gamma Y_{i0} + \beta \sum_j X_{ij0} Y_{j0} + \varepsilon_{i1} \label{master-lmmaker} 
\end{equation}

\noindent so that $\gamma$ represents the auto-time-lag dependence for each unit, and $\beta$ represents the  ``network'' cross-unit-time-lag as mitigated by a connection of strength $X_{ij0}$ in the past. This modelling framework extends beyond the simple case of two time points, as a series of $N-1$ equations can model a system at $N$ time epochs.

Unlike previous dichotomization analyses such as \citet{gelman2009sp}, the effect of dichotomizing the network does not lead to a simple bisection; the past property $Y_{j(t-1)}$ plays a role that cannot be ignored. In particular, if there is any kind of relationship between the network configuration and the previous outcome of interest -- for example, if popular people are also happier than the unpopular, then there is a correlation between in-degree and past property -- it is possible that the thresholding mechanism will distort the relationship in other unexpected ways.

In order to test the effect of dichotomization, the valued networks are simulated and hypothetical nodal attributes are created, which are passed both autoregressively and through the network at each time point. The method is as follows:

\begin{itemize}

\item Generate an autoregressive parameter $\gamma$ and a network parameter $\beta$ from a random distribution (in this case, a positive value well below 1. Choose a variance $\sigma^2$ for the error term $\varepsilon_{it}$.

\item Generate a hypothetical correlation between the indegree of a node $X_{.j}$ and the past property $Y_j$.

\item Given the correlation, choose a mean value for the past property, $\mu_Y$, and generate a past property for each unit, marginally $\mathbf{Y}_{j0} \sim N(\mu_Y, 1)$.

\item Generate the error term to produce the present property and outcome $\mathbf{Y}_1$.

\item At each threshold, compute the conversion factor, $\bar{X}_{high}-\bar{X}_{low}$, that represents the change in unit/dimension from the valued to the binary system.

\item Solve the linear model problem for the true value of $X_{ij}$ as well as at each chosen threshold; that is, determine the estimates of the parameters $\gamma$ and $\beta$ and their respective variances.

\item Compare the estimates of the parameters $\gamma$ and $\beta$, to the underlying true values; in the case of $\beta$, dividing by the unit conversion factor to adjust for the change in scale in the network. 

\end{itemize}

The goal is to then choose the threshold value that best approximate the valued case for the linear model. There are several possibilities that present themselves for a ``best'' approximation, namely the threshold choices that give the smallest mean squared error for the autoregressive parameter $\gamma$ or the cross-unit effect $\beta$, or the best fit to the present-time property $\mathbf{Y}_1$ as determined by $R^2$, which is directly comparable between analyses as the equations are identical except for the form of $\mathbf{X}$.

\begin{figure}
\begin{center}
\includegraphics[width=0.95\linewidth]{\imloc 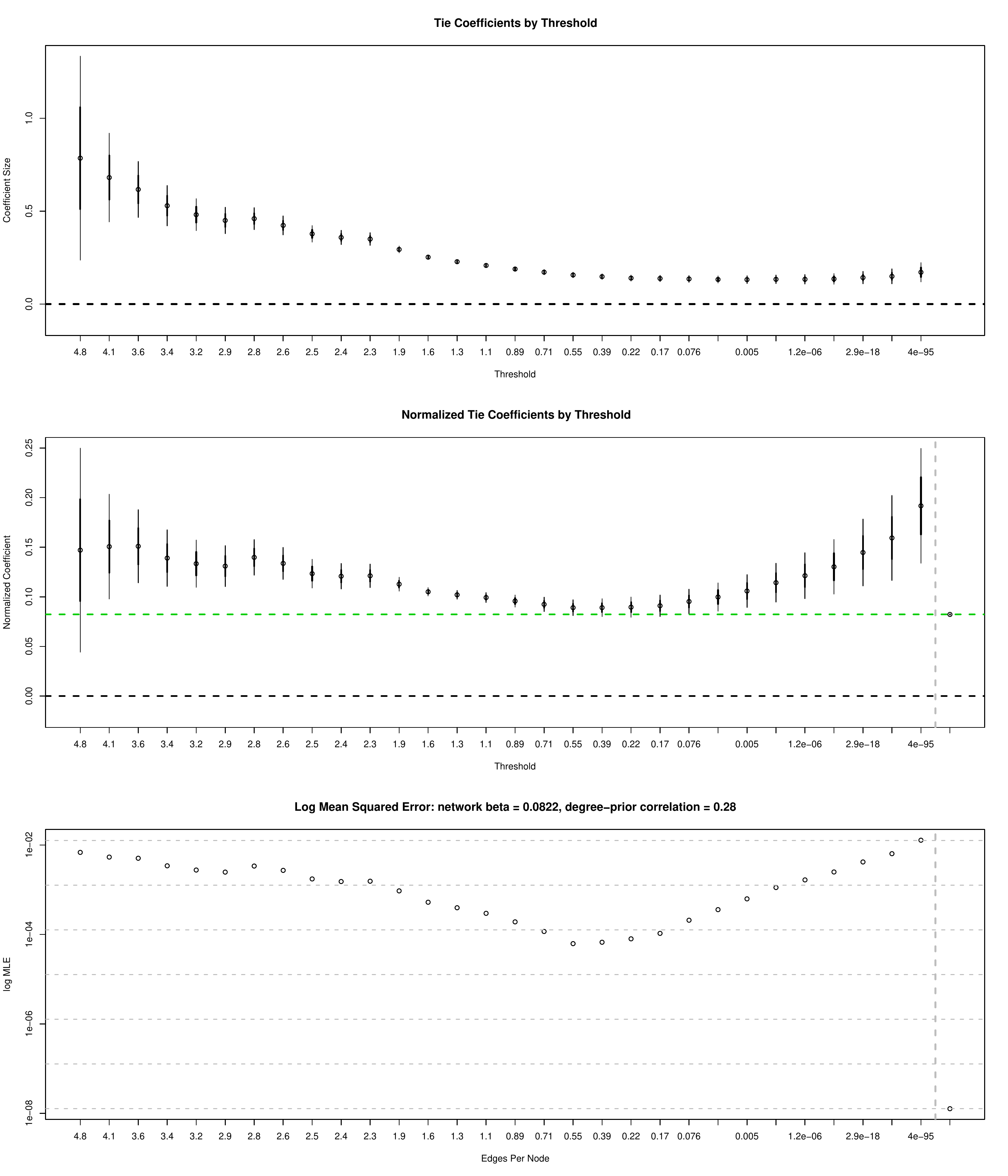}
\end{center}
\caption{\label{thres-single-lm} A comparison of tie coefficients in a linear model across choices of threshold for a fixed underlying toy model. As the choice of threshold increases, the measure of the value of a friendship decreases in absolute terms but remains relatively close to the generation value when accounting for the change in scale. The relative efficiency of the dichotomized models is at least a factor of 1000 below that for the valued model. Contrast with Figure \ref{thres-single-lm-new}.}
\end{figure}

\begin{figure}
\begin{center}
\includegraphics[width=0.95\linewidth]{\imloc 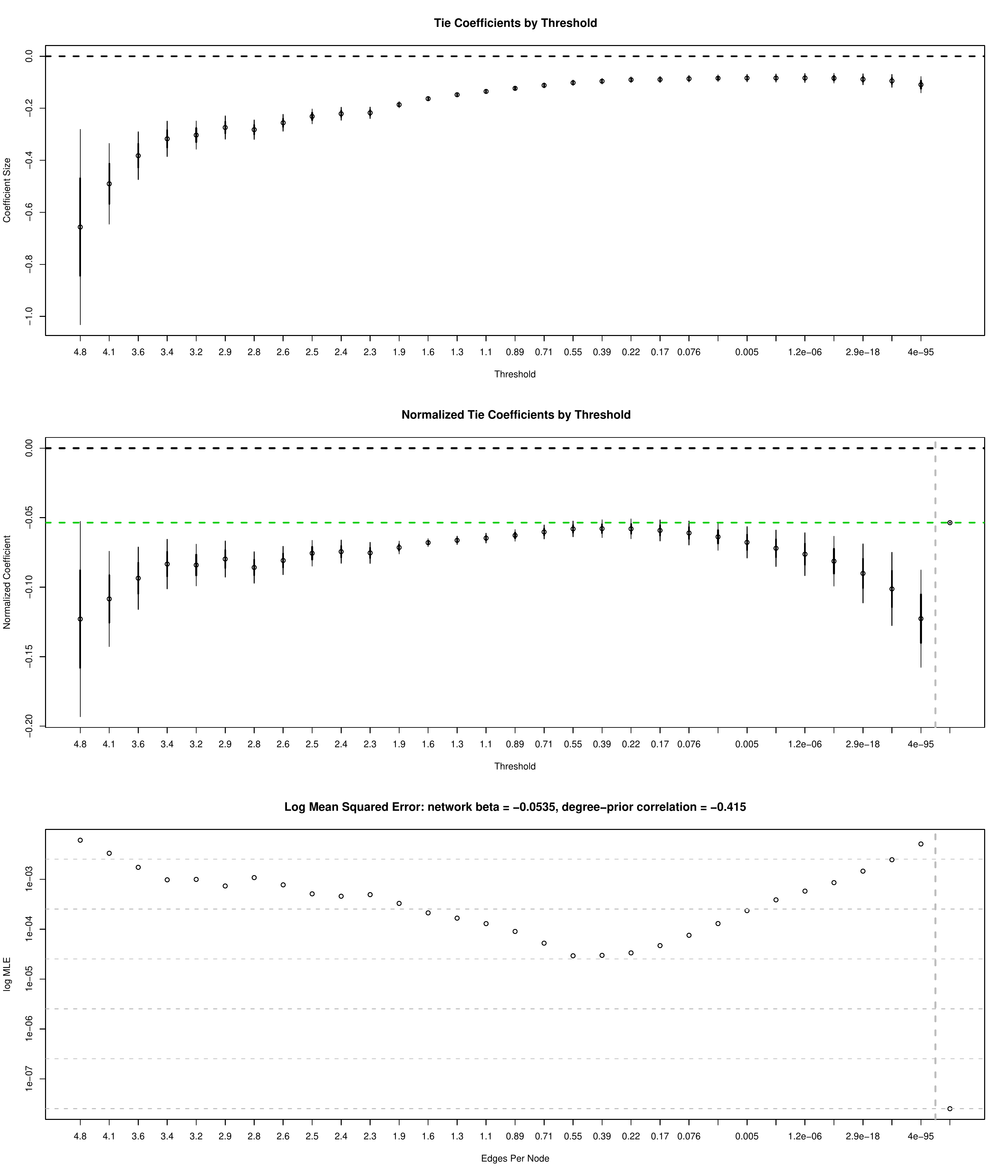}
\end{center}
\caption{\label{thres-single-lm-new} A comparison of tie coefficients in a linear model across choices of threshold for a fixed underlying toy model, using the same underlying network as in Figure \ref{thres-single-lm} but with different generative parameters, including a negative correlation between an individual's indegree. As the choice of threshold increases, the measure of the value of a friendship decreases in absolute terms but remains relatively close to the generation value when accounting for the change in scale. The relative efficiency of the dichotomized models is at least a factor of 1000 below that for the valued model.}
\end{figure}

Two instances of this procedure are demonstrated in Figures \ref{thres-single-lm} and \ref{thres-single-lm-new}, using the same underlying graph structure but using two different underlying time evolutions, wherein the target objective is to minimize the mean squared error of the network effect parameter $\beta$. In Figure \ref{thres-single-lm}, the apparent effect of a tie on a unit's outcome decreases as the number of connections increases, which is to be expected since we are essentially lowering the share of influence that each node has over a target node. However, the 95\% intervals for the scale-adjusted estimates only cover the true parameters in five cases, and the best threshold choice gives a mean squared error nearly 10,000 times greater than the valued model gives for the same analysis.

Figure \ref{thres-single-lm-new} gives a similar pattern, though the signs are flipped as the true beta is negative in this case. The absolute value for $\beta$ is once again inflated with respect to the true value, and the optimal threshold gives a mean squared error roughly 1000 times greater than the valued model analysis.

\subsection{Overall Results: Optimal Cutpoint}

A total of 282 GLM network families were simulated, so that each family has 40 instances of a past-present linear model for a total of 11,280 linear models. For this stage, networks of size 300-600 nodes were added to the analysis, as the analysis of a one-step model is considerably quicker than the geometric and topological decompositions used in Section \ref{threshold-geometry-section}. A summary of simulation inputs, selection criteria and outputs is given in Table \ref{lm-criteria-table}.


\begin{table}
\begin{center}
\begin{tabular}{c|c|c}
Input & Criteria/Condition & Output \\
\hline
Heterogeneity $\sigma_{\alpha}$ & Min $\gamma$ MSE & Edges Per Node (cutpoint)\\
Assortativity $\chi$ & Min $\beta$ MSE  & MSE for $\gamma$ \\
Size $n$ & Max $R^2$ & MSE for $\beta$ \\
Geometry &  & Coverage for $\gamma$ \\
Autocorrelation $\gamma$ &  & Coverage for $\beta$ \\
Network Effect $\beta$ &  & Outcome $R^2$\\
Indegree/Property Correlation $\rho$ &  & \\
Property Mean $\mu$ &  & 
\end{tabular}
\end{center}
\caption{\label{lm-criteria-table} The possible combinations of inputs and outputs for studying the effects of thresholding on linear model parameters. There are 8 possible inputs, 3 optimization conditions and 6 outputs of interest for a total of 144 1-on-1 comparisons.}
\end{table}

\begin{figure}
\begin{center}
\includegraphics[width=0.7\linewidth]{\imloc 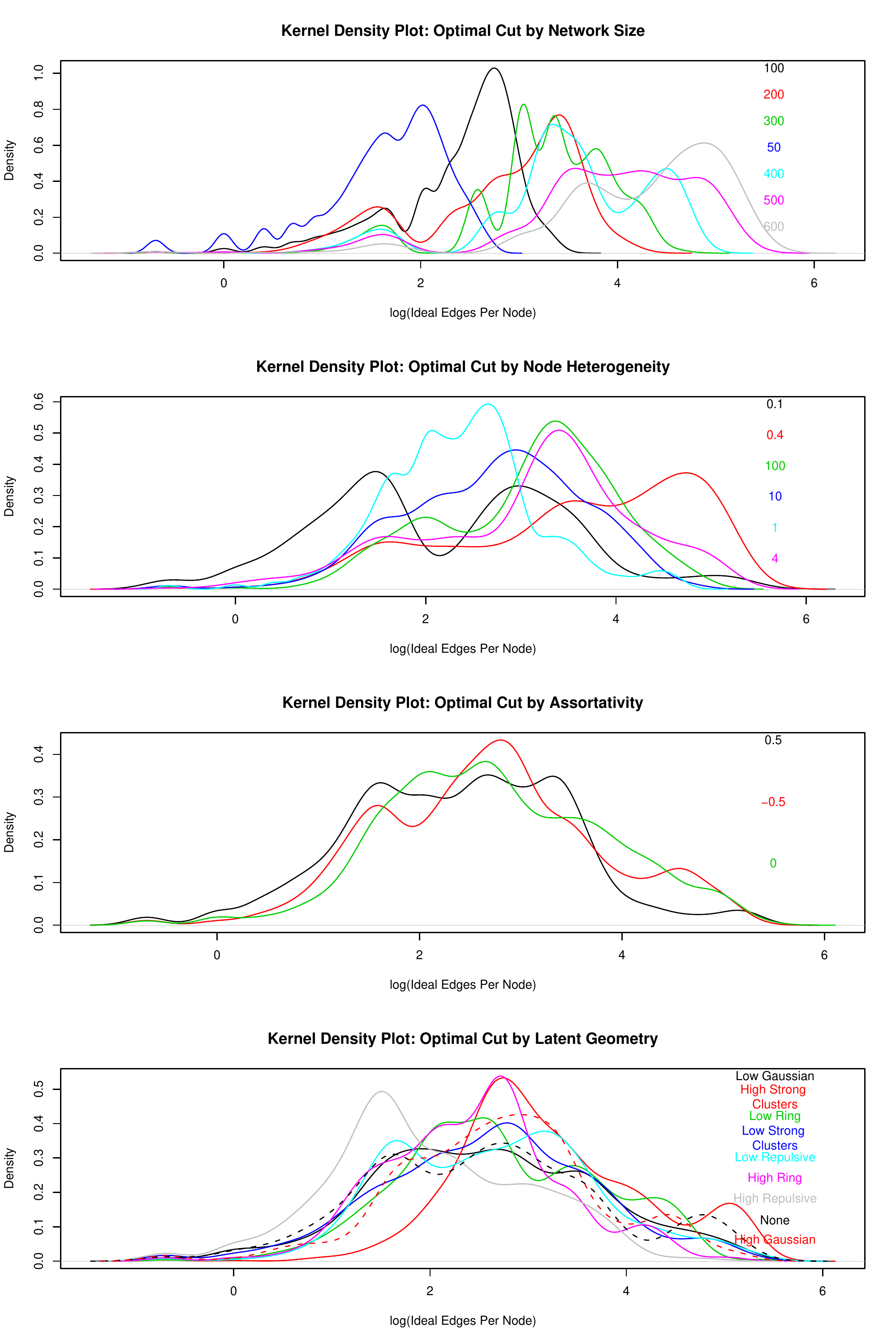}
\end{center}
\caption{\label{lm-ideal-points} Ideal thresholds for linear modelling, as divided by network size, with respect to minimizing the MSE for the network component $\beta$ (results are similar for other criteria.) Network size raises the ideal edges per node (top), but the relative densities in the optimal cases are roughly identical (second from top). There is no apparent linear progression in ideal size by increasing node heterogeneity alone (third), or by (dis)assortative mixing (fourth). Two geometries appear to have a distinct effect on the ideal network size: when nodes are arranged in clusters that prefer external connections (grey), the density may be lower; when nodes are in clusters that prefer internal connections (solid red), a higher density is required.}
\end{figure}

To examine the ideal density as a factor of input properties, Figure \ref{lm-ideal-points} gives a series of kernel density plots for each of the network generation factors; size, heterogeneity, assortativity and geometry. In the top panel, the ideal number of edges per node increases with network size; however, as seen in the second panel, when scaling as a function of network density, or the total number of edges as a fraction of all possible $n(n-1)$ edges, the differences between the network sizes diminishes. There is no apparent linear progression in ideal size by increasing node heterogeneity alone (third), or by (dis)assortative mixing (fourth). 

Of the latent geometries used to simulate the system, two appear to have a distinct effect on the ideal network size: when nodes are arranged in clusters that prefer external connections, the density may be lower; when nodes are in clusters that prefer internal connections, a higher density is required to best approximate the valued system.

A quick inspection of graphs showing various linear model parameter values demonstrated that none of the parameters appreciably affect the optimal cut point.

\subsection{Overall Results: Mean Squared Error Ratio}

One measure of efficiency of an estimator is the mean squared error from the true value, equal to the square of the estimator's bias plus its variance. Given that the true values of the underlying parameters are known, the MSE can be quickly computed for each thresholded value as well as that estimated by the valued model.

\begin{figure}
\begin{center}
\includegraphics[width=0.8\linewidth]{\imloc 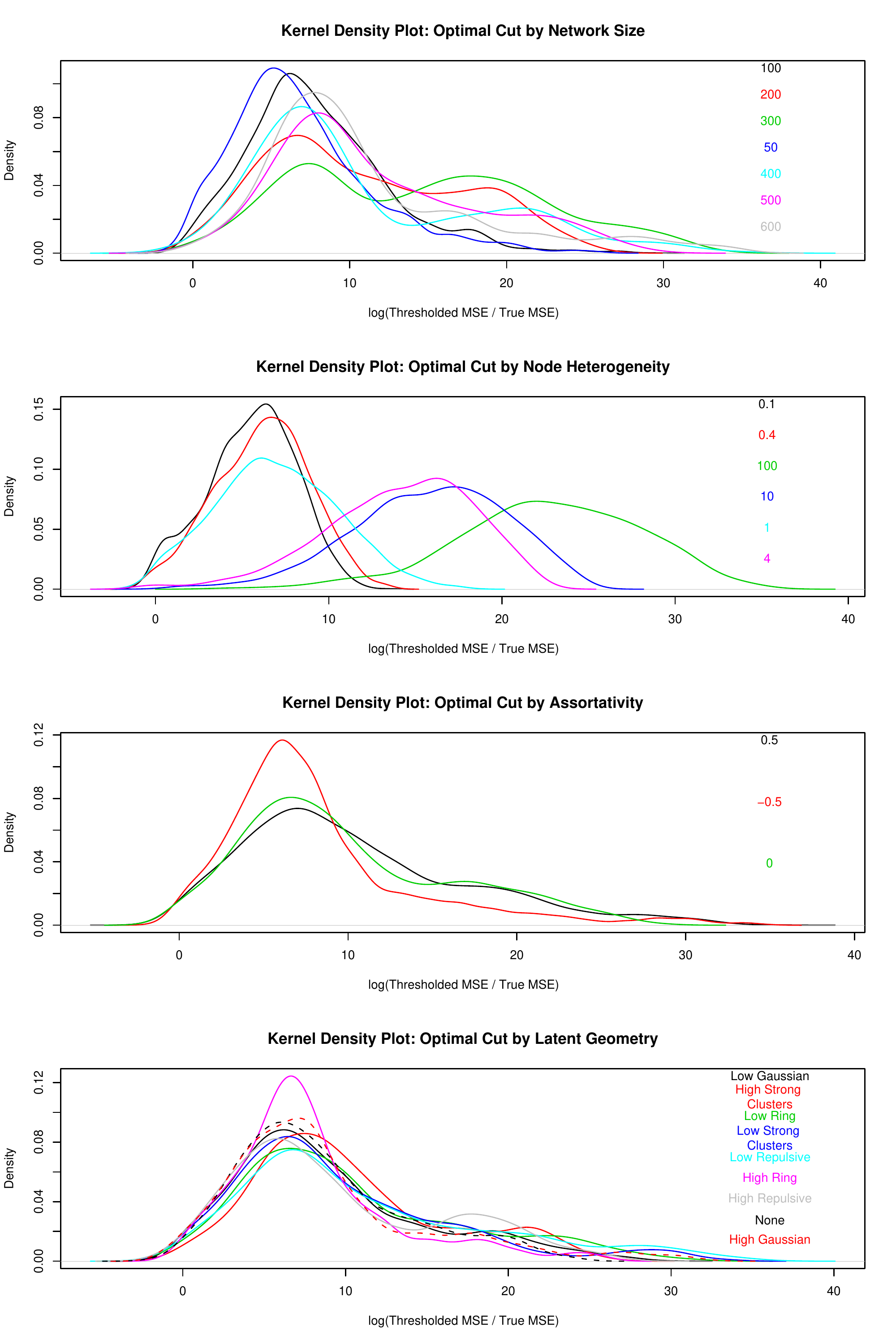}
\end{center}
\caption{\label{all-minmse} Comparing the minimum mean squared error for the estimate of $\beta$ with respect to the underlying valued model. Top: dichotomizing larger networks tends to produce a larger mean squared error in the estimate, though the effect is small compared to the distortion caused by node heterogeneity (second from top) or assortative mixing (third). There is minimal discrimination between graphs due to their latent geometry (bottom).}
\end{figure}

Figure \ref{all-minmse} contains kernel density estimates for the (log) ratio of the optimal threshold for a valued model against the valued model itself, as it varies by generative parameter. First, there is some differentiation between the MSE ratio as broken down by network size, though with the exception of the smaller graphs (50 or 100 nodes), there is no highly suggestive pattern or trend to indicate that larger networks have a higher average MSE than their smaller counterparts.

There is, however, considerable separation between classes of heterogeneity on popularity. As heterogeneity increases between nodes in a network, the optimal MSE ratio rises considerably. Disassortative mixing also appears to lower the optimal MSE ratio. This is likely due to the removal of less popular nodes in the thresholded systems under assortatively mixed systems (where less popular nodes connect to each other, rather than to the system as a whole) or heterogeneous systems (where less popular nodes connect to far fewer nodes in total.)

A quick inspection of graphs showing various linear model parameter values demonstrated that none of the parameters appreciably affect the MSE ratio.

\subsection{Overall Results: Coverage Characteristics}

\begin{figure}
\begin{center}
\includegraphics[width=\linewidth]{\imloc 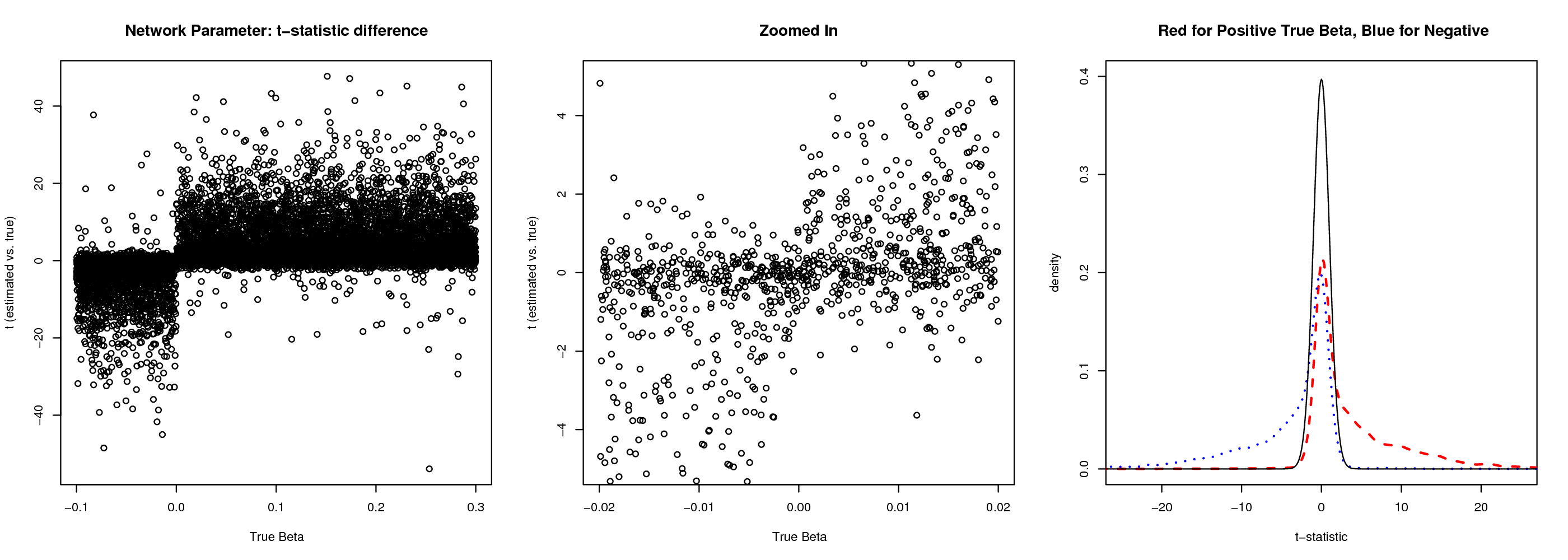}
\end{center}
\caption{\label{lm-beta-split} The differences between the estimated $\beta$ coefficient and the generated true value expressed as t-statistics, after correction for the change in units. There is a sharp change at zero, as seen in closeup in the middle image; for negative true values, the inferences are biased more negatively, and for positive values, the inferences are biased more positively, suggesting strongly that the network coefficient estimate is inflated in magnitude. The rightmost image is a kernel density plot when the true beta is separated by sign, supporting this contention.}
\end{figure}

This section examines the effect that input parameters have on the coverage for estimates of the autoregressive and network effects. One is immediately apparent in Figure \ref{lm-beta-split}, as there is a considerable bias in the estimate of the network effect depending on its sign. Effects are measured to be considerably greater in magnitude than their true values -- more positive in the positive case, more negative in the negative case. Moreover, this effect is invariant in the $t$-statistics with respect to the scale of the coefficient, suggesting that the bias on the coefficient scales with the underlying true value, and is hence a multiplicative effect.

\begin{figure}
\begin{center}
\includegraphics[width=0.9\linewidth]{\imloc 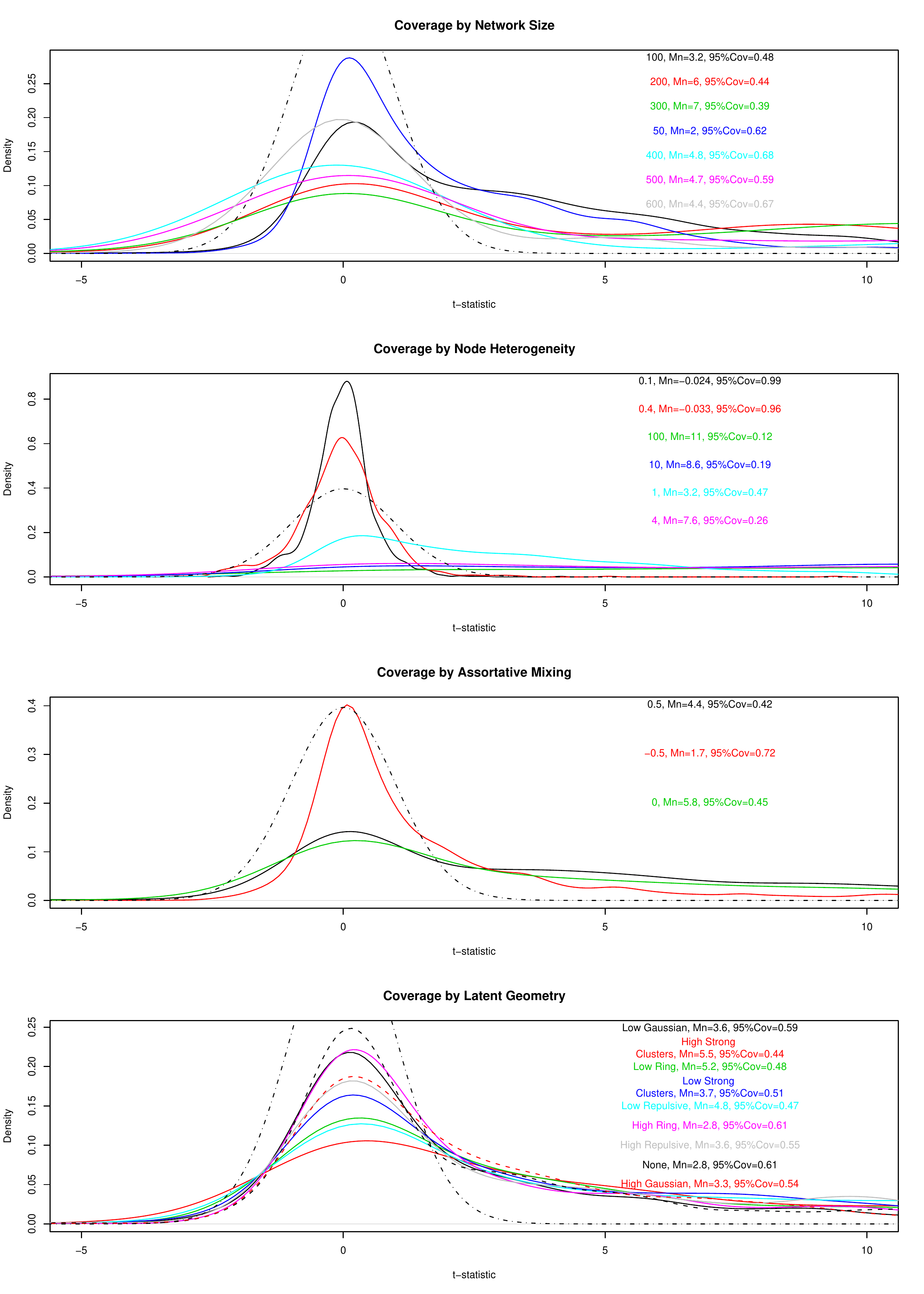}
\end{center}
\caption{\label{lm-beta-split-geo} For only positive values of the true $\beta$, the differences between the estimated coefficient and the generated true value expressed as t-statistics across values of the network generative parameters. The dash-dot line in each plot represents the density of a $t$-distribution with 50 degrees of freedom and is given for reference.}
\end{figure}

The results are then broken down by generative parameter, keeping only positive true values of $\beta$ are included. Using this breakdown, Figure \ref{lm-beta-split-geo} explains much of the additional bias present in the system. While there appear to be differences in the coverage by network size and latent geometry, it is node heterogeneity and assortative mixing that produce the highest biases in coefficient value, in a fashion similar to these variables' effect on network geometry.

The coverage properties of homogeneous network systems are the only ones that are empirically too large, with the central 95\% region of a 50-plus-degree $t$ distribution carrying 99\% and 96\% of the observed outcomes for node heterogeneity of 0.1 and 0.4; all other cases yield dramatically worse coverage. Interestingly, as demonstrated in Figure \ref{lm-ideal-points}, the underlying density at the optimal cut point is not appreciably bigger for highly heterogeneous systems; this suggests instead that the bias is caused by an unequal representation of the most popular nodes in the system driving the response, similar to a standard horizontal outlier having disproportionate leverage over a simple linear model.

There is also a considerable difference in coverage between disassortatively mixed systems and their neutral or assortative counterparts. In the geometric case, the inclusion of disassortative characteristics allowed for a lower density (higher threshold) graph; in the linear model case, the inclusion of links between high- and low-popularity individuals ensures that more units are properly represented in the coefficient estimate, similar to the networks with low heterogeneity in popularity. 

\section{Conclusions}

We have covered a number of different issues with dichotomization, a commonly used method of simplifying a data set by compression into a binary framework. It proves problematic to match the geometric characteristics of a dichotomized graph to the original valued graph, since different statistical summaries are optimized at different threshold values. Estimating the parameters in linear models becomes more problematic, as there is a persistent inflationary bias in the value of the network coefficient when adjusting for the difference in scale using the methods most common in the standard linear model framework, and the probability and confidence intervals produced as a result have remarkably lower coverage than as advertised.

This exploration of the effects of dichotomizing valued data is by definition incomplete. The number of complex network systems appearing in the literature goes far past the range of the synthetic definitions presented herein, and even then, the definition of ``optimal'' is always debatable. By choosing a simple standard -- that the optimal threshold for dichotomizing a valued graph is that which best preserves the features of interest -- it is with some hope that this will contribute to a decrease in the \textit{ad hoc} dichotomization of data purely for convenience.


\subsection{Is dichotomization necessary?}

The motivations for dichotomization in Section \ref{thres-motiv} must be revisited in order to fully appreciate its value as a scientific instrument.

\begin{itemize} 

\item For use in \textbf{exclusively binary methods}, there is little doubt that it is necessary to choose a threshold value in order to put them to use. The appropriateness of shoehorning this data into these models, however, is questionable if the data are already fully formed. For example, the Watts-Strogatz and Barabasi-Albert models are both attempts at producing both evolution stories and replication models for binary systems. The classification of a valued network according to a binary standard may not be particularly useful in light of better quantities already available for valued data.


\item For \textbf{ease of input and data collection}, there remains the risk of error propagation when dichotomization is conducted too early in the investigation. Survey methodology provides better alternatives when it comes to reliable data collection than premature bifurcation.

\item For \textbf{ease of graphical output}, there is no statistical issue to debate; producing informative graphics is a question for aesthetics and the experimenter. The only issue to consider in this case is the distortion of distance; for example, whether it would be better to consider distantly connected nodes as isolates rather than accurately try to demonstrate their positions. 

\item The issue of \textbf{nonlinearity}, particularly in the case of linear modelling, is moot in the case of dichotomization, as it represents only one of a set of transformations that can apply in this case. Additionally, the dramatic loss of efficiency, 100-fold or more in simulations, makes dichotomization extremely unwise unless there truly is a threshold effect in the system to be studied.

\end{itemize}


\subsection{Improving Dimensional Transformation Estimation}

Because the end effect of this sort of network compression is lossy and unpredictably non-linear, the approximation provided by the transformation of units is by no means a perfect mechanism for comparing valued graphs to their dichotomized counterparts. Whether there is a more sophisticated way to compare network effects under data compression schemes is a matter of additional research. However, because the motivations for the procedure do not appear to hold up under scrutiny, the value of such an investigation seems to be purely academic. 

\subsection{Possibilities of Integrated Multiple-Graph Approaches}

It is of course possible to choose a threshold ladder and produce a series of analyses at each threshold choice; this ``multiple slice'' method appears to be a reasonable path to take if a single threshold would be too uncertain. However, the dependence between dichotomized values means that any uncertainties in the estimation process cannot be added as independent quantities, so that losses in efficiency cannot easily be reclaimed by stacking a series of dichotomized networks. For analysis of a system, the problem of combining analyses from a threshold ladder is still open.

For graphical display, there is a reasonable method for using multiple thresholds for graphical purposes, called the ``wedding cake'' model and described in greater detail in the ElectroGraph package for R. The procedure begins by solving for the positions of the nodes in two dimensions for the valued graph, then by sequentially plotting the ties visible at each threshold. In this way, the coordinates for each plot remain the same as each layer of the system is visually examined.\footnote{This act of multiple slicing is similar to what would be known as tomographic analysis. However, the term ``network tomography'' is already in use as a term for inferring the properties of a network by examining its path structure \citep{vardi1996ntestifld}, and so we use the term ``wedding cake'' for its visual interpretation.}

\subsection{Alternative Dichotomization Procedures}

As inspired by the accidental censoring of outbound binary network edges to an upper limit of $k$ \citep{thomas2010cocisnpeaa}, one possibility is the deliberate limitation of outgoing edges corresponding to the $k$-highest valued ties. \citet{thomas2011vttfliwndnebo} show that this performs worse that the standard thresholding criterion in terms of preserving known features of the graph or linear model coefficient estimates.

Keeping the goal of maintaining the inherent structural properties of a valued network in its dichotomized form doesn't require the thresholding tool to perform it. However, it does provide an excellent starting point from which to begin a search of binary graphs that better correspond to their valued counterpoints. A simple simulated annealing procedure, for example, is one in which an edge is added or subtracted from the dichotomized version and compared to the valued graph; the ``energy'' can be expressed as a function of the value or rank discrepancy in the preservation criterion, and edges can be added or subtracted according to a Metropolis-style acceptance procedure until the global minimum is found.



\bibliographystyle{\baseloc ims}
\bibliography{\baseloc actbib}
\end{document}